\documentclass[twocolumn]{aastex631}

\usepackage{amsmath}
\usepackage{amssymb}

\usepackage{graphicx}
\usepackage{subfigure} 

\newcommand       \GBP       {G_{\rm BP}}
\newcommand       \GRP       {G_{\rm RP}}
\newcommand       \mum        {\,{\rm \mu m}}
\newcommand       \Ks           {{ K_{\rm S}}}
\newcommand       \K             {\,{\rm K}}
\newcommand       \Teff           {{ T_{\rm eff}}}
\newcommand       \Angstrom     {\,{\rm \AA}}

\newcommand{\AV}{A_{\rm V}}

\newcommand{\ABP}{A_{G_{\rm BP}}}
\newcommand{\ARP}{A_{G_{\rm RP}}}
\newcommand{\RV}{R_{\rm V}}

\usepackage{booktabs}
\usepackage{longtable}
\usepackage{multirow}

\begin{document}

\title{The Ultraviolet to Mid-infrared Extinction Law of the Taurus Molecular Cloud Based on the Gaia DR3, GALEX, APASS, Pan-STARRS1, 2MASS, and WISE Surveys}

\author{Ling Li}
\affiliation{Department of Astronomy, China West Normal University, Nanchong 637002, China; 
\href{mailto:shuwang}{shuwang@nao.cas.cn},
\href{mailto:xiaodian}{chenxiaodian@nao.cas.cn},
\href{mailto:qingquan}{qqjiangphys@yeah.net}}
%\email{shuwang@nao.cas.cn, chenxiaodian@nao.cas.cn, qqjiangphys@yeah.net}
\affiliation{CAS Key Laboratory of Optical Astronomy, National Astronomical Observatories, 
Chinese Academy of Sciences, Beijing 100101, China}

\author[0000-0003-4489-9794]{Shu Wang}
\affiliation{Department of Astronomy, China West Normal University, Nanchong 637002, China; 
\href{mailto:shuwang}{shuwang@nao.cas.cn},
\href{mailto:xiaodian}{chenxiaodian@nao.cas.cn},
\href{mailto:qingquan}{qqjiangphys@yeah.net}}
\affiliation{CAS Key Laboratory of Optical Astronomy, National Astronomical Observatories, 
Chinese Academy of Sciences, Beijing 100101, China}

\author[0000-0001-7084-0484]{Xiaodian Chen}
\affiliation{Department of Astronomy, China West Normal University, Nanchong 637002, China; 
\href{mailto:shuwang}{shuwang@nao.cas.cn},
\href{mailto:xiaodian}{chenxiaodian@nao.cas.cn},
\href{mailto:qingquan}{qqjiangphys@yeah.net}}
\affiliation{CAS Key Laboratory of Optical Astronomy, National Astronomical Observatories, 
Chinese Academy of Sciences, Beijing 100101, China}
\affiliation{School of Astronomy and Space Science, University of the Chinese Academy of Sciences, Beijing 101408, China} 

\author{QingQuan Jiang}
\affiliation{Department of Astronomy, China West Normal University, Nanchong 637002, China; 
\href{mailto:shuwang}{shuwang@nao.cas.cn},
\href{mailto:xiaodian}{chenxiaodian@nao.cas.cn},
\href{mailto:qingquan}{qqjiangphys@yeah.net}}

\begin{abstract}
Interstellar dust extinction law is essential for interpreting observations. In this work, we investigate the ultraviolet (UV)--mid-infrared (IR) extinction law of the Taurus molecular cloud and its possible variations. We select 504,988 dwarf stars (4200\,K $\leq \Teff \leq$ 8000\,K) and 4,757 giant stars (4200\,K $\leq \Teff \leq$ 5200\,K) based on the stellar parameters of Gaia DR3 as tracers. We establish the $\Teff$--intrinsic color relations and determine the intrinsic color indices and color excesses for different types of stars. In the determination of color excess ratios (CERs), we analyze and correct the curvature of CERs and derive the UV--mid-IR CERs of 16 bands. We consider different effective wavelengths for different types of stars when converting CERs to relative extinction, and obtain the extinction law with a better wavelength resolution. In addition, we analyze the possible regional variation of extinction law and derive the average extinction law of $\RV=3.13\pm0.32$ for the Taurus molecular cloud. Only 0.9\% of subregions have deviations $>3\sigma$, indicating limited regional variation in the extinction law. 
We also discuss the effect of Gaia $\Teff$ overestimation on the determination of the Taurus extinction law and find that the effect is negligible.
\end{abstract}
\keywords{ Interstellar dust extinction (837); Interstellar extinction (841); Interstellar reddening (853); Reddening law (1377)}

\section{Introduction} \label{sec:Introduction}
The wavelength-dependent dust extinction along a sightline, the interstellar extinction curve, is usually expressed as $A_\lambda/\AV$, where $\AV$ is the $V$-band extinction.
The ultraviolet (UV) to optical extinction curves of the Milky Way can be characterized as a one-parameter function of $\RV \equiv \AV/E(B-V)$, where $E(B-V)$ is the color excess \citep{1989ApJ...345..245C}. 
The value of $\RV$ depends on the interstellar environment.
Generally, low-density diffuse regions usually have lower $\RV$ values, while dense regions have higher ones. 
The Galactic diffuse interstellar medium (ISM) sightlines have an average value of $\RV \approx 3.1$\citep{1989ApJ...345..245C, 2003ARA&A..41..241D, 2011ApJ...737..103S, 2019ApJ...877..116W}. 
Lines of sight penetrating into dense environments usually have $4 < \RV < 6$ \citep{1990ARA&A..28...37M}. 
\citet{2013ApJ...771...68P} found that UV extinction at high latitudes is not consistent with the parameterized $\RV$ extinction curves.
\cite{2007ApJ...663..320F} studied the Galactic UV to infrared (IR) interstellar extinction curves with B- and O-type stars and found that 9.5\% of the 328 line-of-sight directions have $\RV>4$. 
In contrast, based on 37,000 stars in the Galactic disk, \cite{2016ApJ...821...78S} found that only 0.8\% of stars have $\RV>4$, and almost no stars with $\RV>4.5$. 
How many regions in the Galaxy have $\RV$ greater than 4, and how large can the $\RV$ value of dense regions be?
To answer these questions, we explore the extinction curve and its possible variations within the Taurus molecular cloud.

The Taurus Molecular Cloud is a nearby star-forming region that often serves as the  laboratory for studying star formation processes \citep{2018AJ....156..271L}. 
Star-forming regions are generally dense regions with accumulated gas and dust. 
Therefore, as a dense star-forming region, the Taurus molecular cloud may have a larger $\RV$ value. 
In addition, the complex structure can be clearly seen in the CO emission map of the Taurus molecular cloud, indicating that the Taurus molecular cloud has a complex interstellar environment, which is very suitable for analyzing the variation of extinction law with interstellar environments. 

In recent years, there has been an increased interest in extinction laws. Many works have studied the extinction laws in different regions of the Milky Way, such as the Galactic plane, the Galactic center, and the high Galactic latitudes \citep{2014A&A...561A.142T, 2016ApJ...821...78S, 2018ApJ...859..137C, 2019ApJ...886..108F, 2020ApJ...895...38H, 2020MNRAS.496.4951M, 2020ApJ...891...67M, 2022ApJ...930...15D, 2022MNRAS.514.2407S, 2023arXiv230401991G}. Similarly, extinction studies have also been performed on nearby galaxies such as the Magellanic Clouds, M31, and M33 \citep{2021ApJ...922..135D, 2022AJ....163...16M, 2022ApJS..259...12W, 2022ApJS..260...41W, 2023A&A...671L..14F, 2023ApJ...946...43W}.

To explore the UV--IR extinction law and its possible variations within the Taurus molecular cloud, we adopt dwarf stars and giant stars as tracers. These stars are selected based on stellar parameters from Gaia DR3 \citep{2022arXiv220800211G} catalog. We gather photometric data from several survey projects, including GALEX \citep[Galaxy Evolution Explorer,][]{2005ApJ...619L...1M}, APASS \citep[The American Association of Variable Star Observers Photometric All-Sky Survey,][]{2014CoSka..43..518H}, Pan-STARRS1 \citep[PS1,][]{2004AN....325..636H}, 2MASS \citep[Two Micron All-Sky Survey,][]{2006AJ....131.1163S}, and WISE \citep[Wide-field Infrared Survey Explorer,][]{2010AJ....140.1868W}.
We establish the effective temperature $T_{\rm eff}$--intrinsic color relations of giants and dwarfs to obtain intrinsic colors. For each star, we calculate the color excess (CE). Then the color excess ratios (CERs) are derived by the statistical CE method.
After that, we convert the CERs into the relative extinction $A_\lambda/\ARP$ and obtain the UV to IR dust extinction curves of the Taurus molecular cloud. 
Finally, we analyze the variation of the extinction law across the Taurus region. In addition, we explore whether the extinction law is consistent across different stellar measurements. We try to establish a standardized procedure for deriving the extinction law.
The structure of this paper is as follows. In Section~\ref{sec:Data and Sample}, we present data sets and the selection of samples. In Section~\ref{sec:method}, we describe the methods to derive the relative extinction. We report in Section~\ref{Result and Discussion} the derived UV to mid-IR relative extinction $A_\lambda/\ARP$, as well as the $\RV$-dependent extinction law. We also discuss the variation of extinction law within the Taurus region in Section~\ref{Result and Discussion}. We summarize our main conclusions in Section~\ref{Summary}.

\section{Data and Sample} \label{sec:Data and Sample}
\subsection{Data}\label{sec:Data}

The Taurus molecular cloud is a remarkable dark cloud centered at $l \sim 170^\circ$ and $b \sim -15^\circ$, and covers an area of about 900 deg$^2$ \citep{2001ApJ...547..792D}.
The distance to the Taurus cloud is estimated to be $\sim 140$\,pc \citep{2001ApJ...547..792D}, $\sim 126.6$ to 162.7\,pc \citep{2018ApJ...859...33G}, $\sim 128.5$ to 198.1\,pc \citep{2019A&A...630A.137G}, $\sim 135 \pm 20$ pc \citep{2014ApJ...786...29S}, $\sim 145_{-16}^{+12}$\,pc \citep{2019A&A...624A...6Y}, and $\sim 141 \pm 2 \pm 7$\,pc \citep{2019ApJ...879..125Z}.
Therefore, we only use stars within 1 kpc to study the extinction law of the Taurus region, which can effectively avoid the contribution of background clouds to the extinction. 
We determine the specific study area of the Taurus region based on the CO emission map of \citet{2001ApJ...547..792D}. 
The final selected Taurus region covers $150^\circ\le l \le190^\circ$ and $-29^\circ\le b \le-2^\circ$. 
We construct giant and dwarf samples with stellar parameters from Gaia and collect UV to mid-IR photometric data from the GALEX, Gaia, APASS, PS1, 2MASS, and WISE surveys.

Gaia DR3 is the third data release of the Gaia mission and contains high-precision photometry in three broad bands, $G$, $\GBP$, and $\GRP$ \citep{2022arXiv220800211G}. 
The $G$ band covers the whole optical wavelength ranging from 330 to 1050 nm, while $\GBP$ band and $\GRP$ band cover the wavelength ranges of 330--680 nm and 630--1050 nm, respectively \citep{2018A&A...616A...4E}. 
The Gaia DR3 catalog also provides GSP-Phot (General Stellar Parameterizer from Photometry) results of stellar parameters, such as effective temperature $\Teff$, surface gravity $\log g$, and metallicity [M/H], for 471 million sources \citep{2022arXiv220606138A}. The typical differences of GSP-Phot $\Teff$ and $\log g$ to literature values are 110\,K, 0.2--0.25 \citep{2022arXiv220606138A}.

\begin{figure}[htbp]
\centering
\includegraphics[scale=0.35]{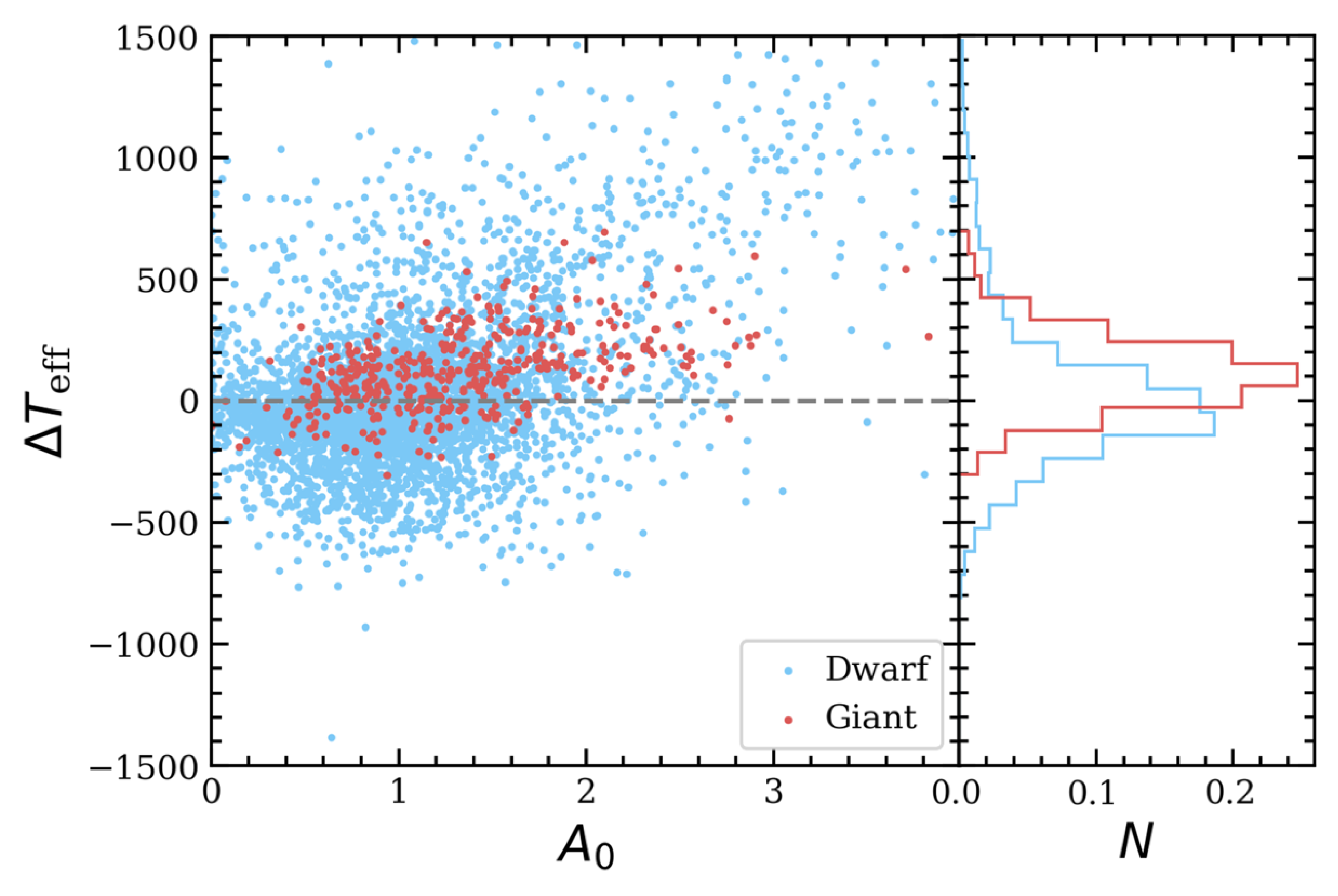}
\caption{Left: Differences in effective temperature $\Delta \Teff = (\Teff)_{\rm Gaia} - (\Teff)_{\rm APOGEE}$ as function of estimated extinction $A_0$. The red dots and blue dots are giants and dwarfs, respectively. The gray dashed line is $y = 0$. Right: Histogram of $\Delta \Teff$ for giants (red) and dwarfs (blue). $N$ is the normalised number.}
\label{fig:APOGaia}
\end{figure}

We compare Gaia $\Teff$ estimates to those from the APOGEE to analyze the reliability of Gaia $\Teff$. 
In the Taurus region, there are 4,554 sources with both Gaia $\Teff$ and APOGEE $\Teff$.  
Figure~\ref{fig:APOGaia} displays the difference in effective temperature $\Delta \Teff = (\Teff)_{\rm Gaia} - (\Teff)_{\rm APOGEE}$ varies with the extinction at 541.4\,nm $A_0$. The red and blue dots are the giants and dwarfs, respectively. It is clear that the overestimation of Gaia $\Teff$ becomes more significant as the extinction $A_0$ increases. The overall trend is the same as that of \cite{2022arXiv220606138A} and \cite{2023ApJS..267....8A}.
\cite{2022arXiv220606138A} indicated that the Gaia $\Teff$ overestimation can reach $1000\K$ at $A_0\sim6$ mag. They also found that at $A_0<4$ mag, the $\Teff$ of 84\% of the total Gaia sources differ from that of APOGEE by less than $1000\K$, and thus the Gaia $\Teff$ is reliable in this regime. In our sample of the Taurus region (Section~\ref{sec:The Giant and Dwarf Samples}), 99\% of the source have $A_0<4$ mag. The mean absolute difference is $\sim125\K$ for the giants and $\sim43 \K$ for the dwarfs, of which only 2.5\% of the sources have a difference greater than $1000\K$. Therefore, the $\Teff$ we use is relatively reliable. Further corrections for the Gaia $\Teff$ deviation are discussed in Section~\ref{sec:comp Gaia}.

The GALEX is the first sky-wide UV survey and provides photometry of two broad bands, far-UV (FUV, $1344-1786\Angstrom$) and near-UV (NUV, $1771-2831\Angstrom$) \citep{2014Ap&SS.354..103B}.
We use data from the catalog of GUVcat\_AIS GR6+7 \citep{2017ApJS..230...24B}.

The APASS photometric survey covers the whole sky and provides a catalog in five filters (Johnson $B,V$ and Sloan $g',r',i'$) for stars in the range of $10\le V \le 17$ mag \citep{2016yCat.2336....0H}. 
As we also collect $g, r, i$-bands photometric data from the PS1 survey, we only use the $B$- and $V$-band photometric data from the APASS DR9 catalog.  

The PS1 survey observed the entire sky north of declination $-30^\circ$ \citep{2004AN....325..636H}. It provides the photometric data in five bands: $g, r, i, z$, and $y$ bands, covering about 0.4--1$\mum$ \citep{2010ApJS..191..376S}.

The 2MASS is a whole-sky survey in the near-IR bands \citep{2003AJ....126.1090C}. We use $J$, $H$, and $\Ks$ photometric data from the 2MASS point-source catalog \citep{2006AJ....131.1163S}. 

The WISE is a mid-IR full-sky survey in four bands: $W1, W2, W3,$ and $W4$ bands with wavelengths centered at 3.35, 4.60, 11.56, and 22.09\,$\mum$, respectively \citep{2010AJ....140.1868W}. We collect WISE $W1, W2,$ and $W3$ bands data from the ALLWISE catalog.

\subsection{The Giant and Dwarf Samples}\label{sec:The Giant and Dwarf Samples}

We adopt giants and dwarfs as extinction tracers to investigate the extinction law.
The giants and dwarfs are selected based on stellar parameters from Gaia GSP-Phot results.
The Gaia GSP-Phot stellar parameters are estimated from the low-resolution BP and RP spectra \citep{2022arXiv220606138A}. 
\cite{2022arXiv220605864C} and \cite{2022arXiv220605992F} further validated GSP-Phot results.
For $\Teff$, the median absolute error is 119\,K, and the mean absolute error is 180\,K. 
For $\log g$, the median absolute error is 0.2\,dex. 
For ${\rm[M/H]}$, the GSP-Phot estimates are systematically underestimated by 0.2 dex. 
In this work, we limit $(\Teff){\rm err} < 180$\,K, $(\log g){\rm err} < 0.2$, and $({\rm[M/H]}){\rm err} < 0.5$\,dex.  

Further, we select giant and dwarf candidates in the $\Teff$--$\log g$ diagram.
The giant sample includes stars with 1 $\leq \log g \leq $ 3.3 and 4200\,K $\leq T_{\rm eff}\leq $ 5200\,K, while the dwarf sample includes stars with $\log g \geq 4$ and 4200\,K $\leq \Teff \leq$ 8000\,K. 
Most of the selected giants have metallicities in the range of  -0.5 $\leq {\rm[M/H]} \leq$ 0.2, while most of the selected dwarfs have metallicities in the range of -1.0 $\leq {\rm[M/H]} \leq$ 0.5. 
Finally, the giant sample contains 4,757 stars, and the dwarf sample contains 504,988 stars, which are listed in Table~\ref{tab:sample}.

By cross-matching these samples with photometric catalogs listed in Section \ref{sec:Data}, the multiband photometric data for the giant sample and the dwarf sample are obtained. To guarantee photometric precision, we select stars that satisfy the following criteria for each photometric catalog:  
\begin{enumerate}
\item For GALEX data, we select stars with photometric error $\leq 0.3$\,mag in NUV band, and no restriction is set on FUV band because of the fewer sources. We use 13.85 mag and 13.73 mag to be the brightward limit magnitude in NUV and FUV bands. 
\item For Gaia data, we select stars with photometric error $\leq 0.02$\,mag and magnitude $\leq 18.0$\,mag in $G$, $\GBP$, and $\GRP$ bands.
\item For APASS data, we select stars with photometric error $\leq 0.05$\,mag in $B$ and $V$ bands.
\item For PS1 data, we select stars with photometric error $\leq 0.05$\,mag in $g, r, i, z$, and $y$ bands. We use 14.5, 15.0, 15.0, 14.0, and 13.0 mag to be the brightward limit magnitudes and 22.0, 21.8, 21.5, 20.9, and 19.7 mag \citep{2016arXiv161205560C} to be the faintward limit magnitudes in $g, r, i, z$, and $y$ bands, respectively.
\item For 2MASS data, we select stars with photometric error $\leq 0.05$\,mag and magnitude ranging from 6.0 to 14.0\,mag in $J, H$, and $\Ks$ bands.
\item For WISE data, we select stars with photometric error $\leq 0.2$\,mag in $W1, W2$, and $W3$ bands.
\end{enumerate}

\section{Method} \label{sec:method}
\subsection{Intrinsic Color Index}\label{sec:intrinsic color}

We adopt the blue-edge method to derive the intrinsic color index of stars. For a set of stars with given stellar parameters, the blue-edge method assumes that the bluest stars are zero-reddening stars, so that their observed colors can represent the intrinsic colors \citep{2001ApJ...558..309D}. \cite{2014ApJ...788L..12W} developed this method and established the relationship between intrinsic color index and $\Teff$ for K-type giants. 
Further, this method is applied to determine the multiband intrinsic colors of different types of stars \citep{2017AJ....153....5J, 2017ApJ...848..106W, 2018ApJ...861..153S, 2023ApJ...946...43W, 2023ApJ...945..132C}. 
The blue-edge method relies on stars with very low extinction, which have more reliable Gaia parameters, and thus this method is almost unaffected by Gaia $\Teff$ overestimation. 
Since the average extinction of the Taurus region is $\sim 4$\,mag in the $V$-band \citep{2005PASJ...57S...1D},  
it may be difficult to find zero-reddening stars in the sightlines towards the Taurus region. Therefore, we selected a diffuse region covering $235^\circ\le l \le245^\circ$ and $-5^\circ\le b \le5^\circ$ to establish the relationship between $\Teff$ and intrinsic colors.  
The specific procedure is as follows.
\begin{enumerate}
\item Considering the effects of ${\rm [M/H]}$ on intrinsic colors, especially in the short-wavelength bands, we divide the dwarf sample into three subsamples according to ${\rm [M/H]}$: $-1 \leq {\rm[M/H]} < -0.5$, $-0.5 \leq {\rm[M/H]} < 0$ and $0 \leq {\rm[M/H]} \leq 0.5$. 
As the [M/H] distribution of the giant sample is very narrow, $-0.5 \leq {\rm[M/H]} \leq 0.2$, we did not divide any subsamples.
\item We select the bluest 5\% of stars in the bin of $\triangle \Teff = 100$ K (central value varying 50 K at each time) on the $\Teff$ vs. 2MASS $(J-\Ks)$, Gaia $(\GBP-\GRP)$ diagrams. These stars are considered as zero-reddening stars. The median $G$-band extinction (given by the Gaia catalog) of zero-reddening stars is 0.043 mag.
\item Using a cubic polynomial function, we fit the zero-reddening stars and establish the relations of $\Teff-(\lambda_1-\GRP)_0$, $\Teff-(\GBP-\lambda_2)_0$. ${\lambda}_1$ are $G$ band from Gaia, $B$ and $V$ bands from APASS, $g$ and $r$ bands from PS1, $J, H,$ and $\Ks$ bands from 2MASS, and $W1, W2$, and $W3$ bands from WISE. ${\lambda}_2$ are $i, z$, and $y$ bands from PS1.
\end{enumerate}

\begin{figure}[htbp]
\centering
\includegraphics[scale=0.25]{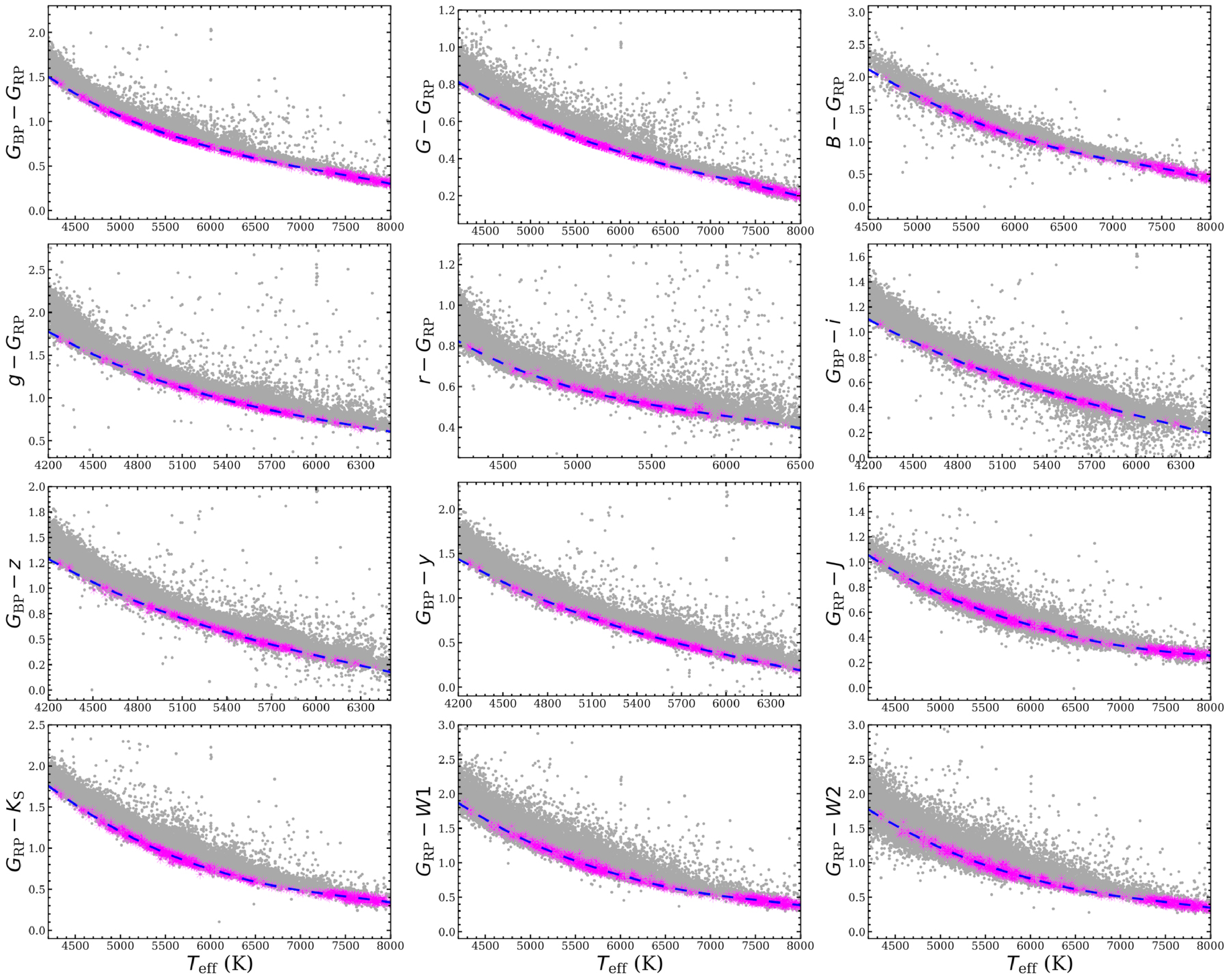}
\caption{Determination of intrinsic color indices based on the blue-edge method. Gray dots are dwarf stars with $-0.5\leq{\rm[M/H]} < 0$ in the diffuse region ($235^\circ\le l \le245^\circ$, $-5^\circ\le b \le5^\circ$) in the $\Teff$ vs. observed color diagrams. Magenta asterisks are the zero-reddening stars selected according to the $\Teff$ vs. $(\GBP-\GRP)$ and $(J-\Ks)$ diagrams. These zero-reddening stars are fitted by a cubic polynomial and the blue dashed lines are the best fitting lines representing the $\Teff$--intrinsic color relations.}
\label{fig:dwarf2 int}
\end{figure}

Figure~\ref{fig:dwarf2 int} shows the $T_{\rm eff}$ vs. observed color diagrams for the diffuse region ($235^\circ\le l \le245^\circ$, $-5^\circ\le b \le5^\circ$) dwarfs with $-0.5\leq{\rm[M/H]} < 0$. The gray dots are all dwarf stars, and the selected zero-reddening stars are magenta asterisks. We obtain the $\Teff$--$(\lambda_1-\GRP)_0$, $\Teff$--$(\GBP-\lambda_2)_0$ relations as the blue dotted lines by fitting these selected stars with a cubic polynomial function.

To get the intrinsic color indices in the GALEX FUV and NUV bands, we adopt the intrinsic color indices and $\Teff$ and metallicity relations given by \cite{2021ApJS..254...38S}. Then, with the $\Teff$ and ${\rm [M/H]}$ of Gaia DR3, we obtain the $({\rm FUV}-\GBP)_0$ and $({\rm NUV}-\GBP)_0$ values of dwarf stars.

\subsection{Color Excess Ratio}\label{sec:Color Excess Ratio}

\begin{figure}[htbp]
\centering
\includegraphics[scale=0.18]{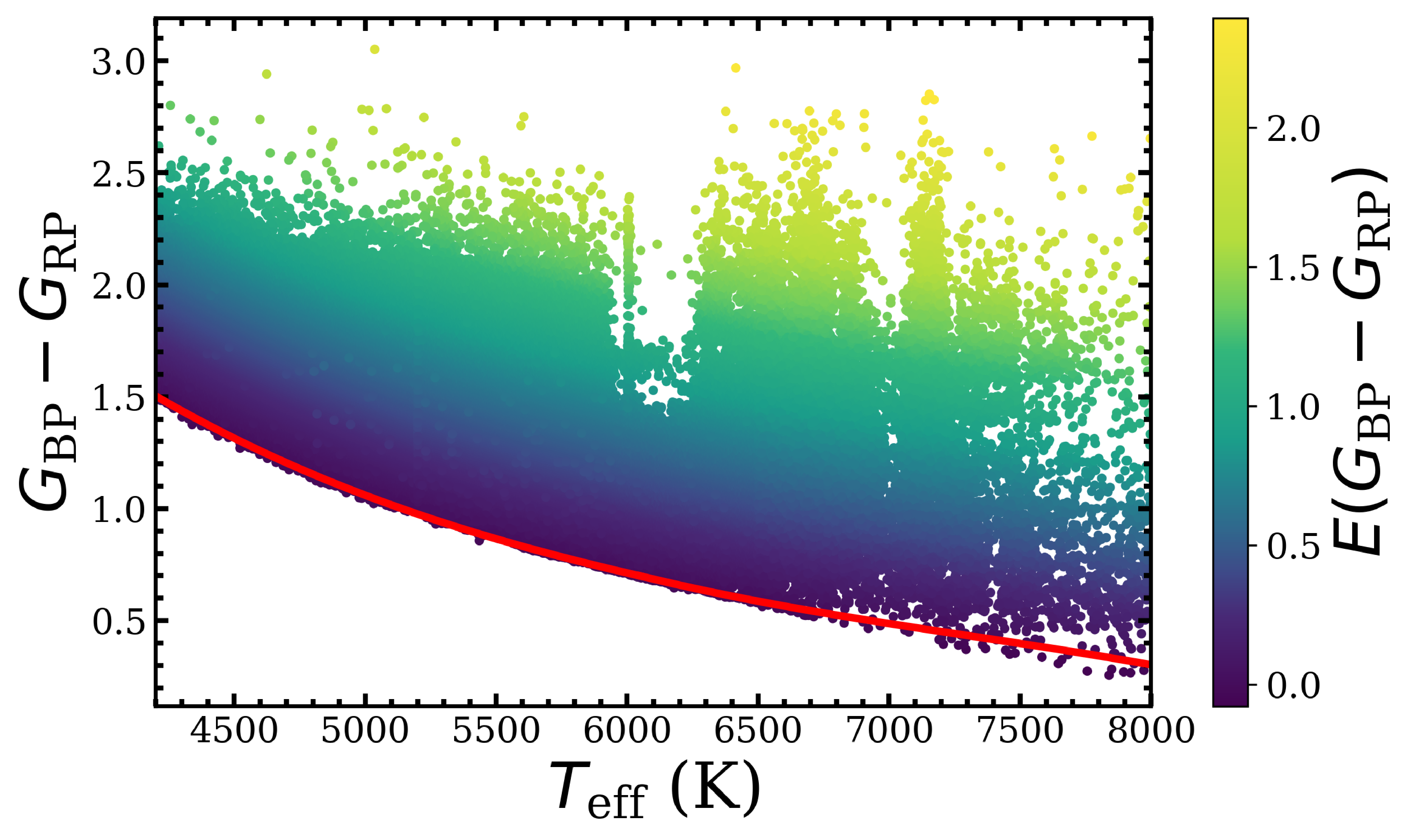}
\caption{$\Teff$ vs. observed color $(\GBP-\GRP)$ diagram for the dwarfs with $-0.5\leq{\rm[M/H]} < 0$ in the Taurus region. The red solid line represents the $\Teff$--$(\GBP-\GRP)_0$ relation, while colors denote the reddening $E(\GBP-\GRP)$ of stars.}
\label{fig:int+CE}
\end{figure}

The CEs are calculated by subtracting the intrinsic color indices from the observed color indices. Figure~\ref{fig:int+CE} is $\Teff$ vs. $(\GBP-\GRP)$ diagram of dwarfs with $-0.5\leq{\rm[M/H]} < 0$ in the Taurus region. 
The color denotes the value of the $E(\GBP-\GRP)$. 
The red solid line represents the intrinsic color. 
Obviously, the closer to the red solid line, the smaller the $E(\GBP-\GRP)$.

We use the CE method to obtain the CERs. This method calculates the ratio of two CEs.
\cite{2019ApJ...877..116W} suggested using high photometric quality bands, such as $\GBP$ and $\GRP$, as the basis bands in the CER analysis to reduce the error caused by the fitting method.
Therefore, we adopt $\GBP$ and $\GRP$ as the basis bands in this work and perform a linear fit to the CE--CE plots to obtain the CERs.
The expression is as follows
\begin{equation}\label{equ:k12}
\begin{aligned}
   k_{{\lambda}_1}=E({\lambda}_1-\GRP)/E (\GBP-\GRP),\\
   k_{{\lambda}_2}=E({\lambda}_2-\GBP)/E (\GBP-\GRP),
\end{aligned}
\end{equation}
where ${\lambda}_1$ are $G$ band from Gaia, $B$ and $V$ bands from APASS, $g$ and $r$ bands from PS1, $J, H,$ and $\Ks$ bands from 2MASS, and $W1, W2$, and $W3$ bands from WISE. ${\lambda}_2$ are FUV and NUV bands from GALEX, and $i, z$, and $y$ bands from PS1.

\begin{figure}[htbp]
\centering
\includegraphics[scale=0.35]{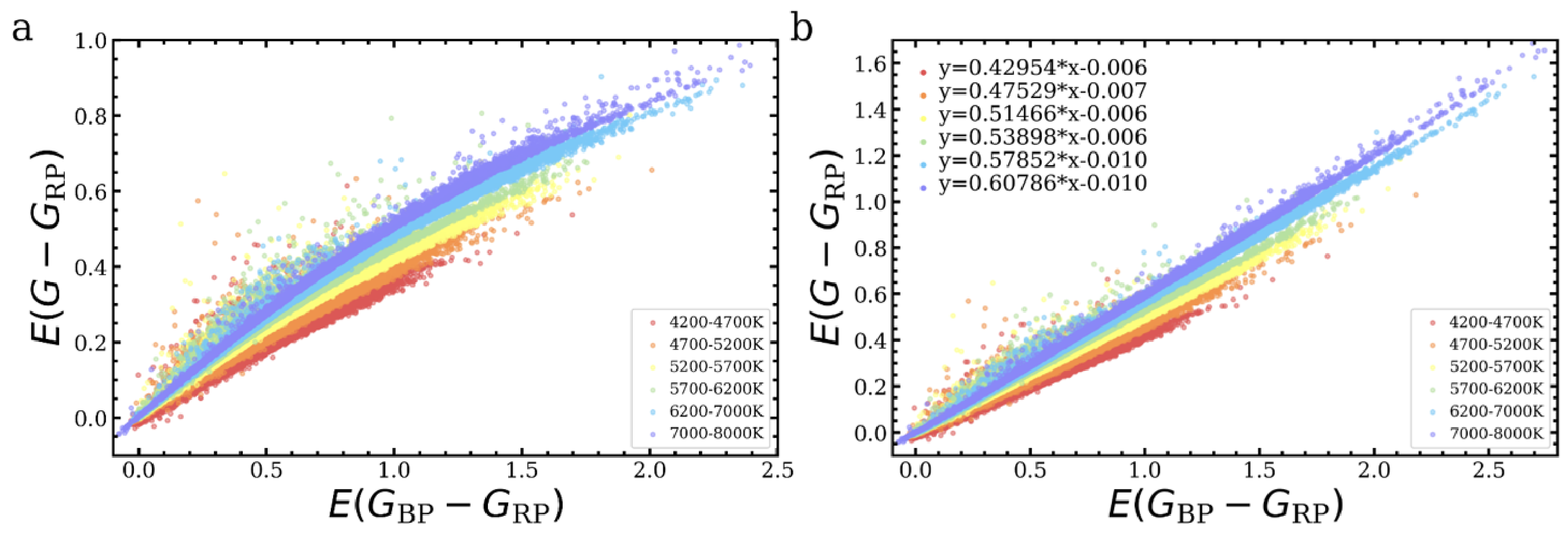}
\caption{The $E(\GBP-\GRP)$ vs. $E(G-\GRP)$ diagrams of the dwarfs with $-0.5\leq{\rm[M/H]} < 0$ before (a) and after (b) the curvature correction. Different colors represent dwarf stars in different $\Teff$ intervals. The results of the linear fitting of dwarfs in different $\Teff$ intervals are listed in the upper left corner of (b).}
\label{fig:ab}
\end{figure}

Figure~\ref{fig:ab} is an example of the CE--CE diagram in the Gaia bands. Different colors represent dwarf stars in different $\Teff$ intervals. 
As shown in Figure~\ref{fig:ab} (a), the curvature of the CER is obvious. This is because when the bandwidth is not infinitely narrow, the effective wavelength of the band becomes longer as the extinction increases. 
In this work, we use the Gaia $\GBP$ and $\GRP$ bands with excellent photometric quality as the basis bands; however, their broad bandwidth may result in curvature of the CE--CE plot. 
In addition, the extent of curvature is related to the spectral type (indicated by the $\Teff$). The higher the $\Teff$, the more severe the curvature phenomenon.

\cite{2019ApJ...877..116W} analyzed the systematic curvature of CERs, and \cite{2023ApJS..264...14Z} also discussed the variation of CERs with $T_{\rm eff}$. 
We use the method of \cite{2019ApJ...877..116W} to analyze and correct the curvature. Besides, we analyze the effect of spectral type on curvature by dividing the dwarf sample according to a two-dimensional grid of [M/H] and $\Teff$.  
The specific procedure is as follows. We first estimate the evolving filter wavelength extinction $A_{\lambda}$ and establish the relation between $A_{\lambda}$ and $A_{\lambda,0}$ (the static wavelength extinction). 
The evolving filter wavelength extinction is calculated by 
\begin{equation}  \label{equ_cur}
A_\lambda=-2.5 \times \log 
\left (\frac{\int F_\lambda(\lambda)S(\lambda)R(\lambda)d\lambda} 
{\int F_\lambda(\lambda)S(\lambda)d\lambda} \right )~~,   
\end{equation}
where $F_\lambda(\lambda)$ is the stellar intrinsic flux, and $S(\lambda)$ is the filter transmission curve. 
$R(\lambda)$ is the wavelength-dependent extinction factor, which is determined by the combination of the static wavelength extinction $A_{\lambda,0}$ and the extinction law.

Specifically, for each dwarf subsample with different ${\rm[M/H]}$ interval, we divide it by a temperature bin of 500 K. For sources with high $\Teff$ ($>6000$ K), we extend the $\Teff$ interval according to the number of sources. The specific classification criteria are listed in Table \ref{tab:sample}. 
As the giant sample covers a relatively narrow range of $\Teff$, we did not divide it further. 
We calculated the average values of $\Teff$, $\log g$, and ${\rm [M/H]}$ for the stars in each [M/H] and $\Teff$ grid. Based on these parameters, we select the corresponding synthetic stellar spectra of \cite{1997A&AS..125..229L} and calculate the $A_{\lambda}$ by Equation~(\ref{equ_cur}).
Then, we calculate the extinction difference between the evolving filter wavelength extinction and the static wavelength extinction, $\triangle A_{\lambda} = A_{\lambda}-A_{\lambda,0}$. 
After that, a polynomial fitting is adopted to analyze the variation of $\triangle A_{\lambda}$ with $E(\GBP-\GRP)$ and to determine the $E(\GBP-\GRP)$--$\triangle A_{\lambda}$ relations. 
Finally, we use these relations to correct the curvature in each band.

Figure~\ref{fig:ab} is an example of the CE--CE diagram $E(G - \GBP)$ versus $E(\GBP - \GRP)$ of dwarfs with $-0.5 \leq {\rm[M/H]} < 0$ before and after the curvature correction.  
It is clear that in Figure~\ref{fig:ab} (a), the heavily reddened stars have larger curvature extent, and stars in different $\Teff$ ranges exhibit different extents of curvature. 
After the curvature correction, the star distribution in Figure~\ref{fig:ab} (b) exhibits good linearity. The linear fitting results are also listed in the upper left corner of Figure~\ref{fig:ab} (b).

\begin{figure}[htbp]
\centering
\includegraphics[scale=0.25]{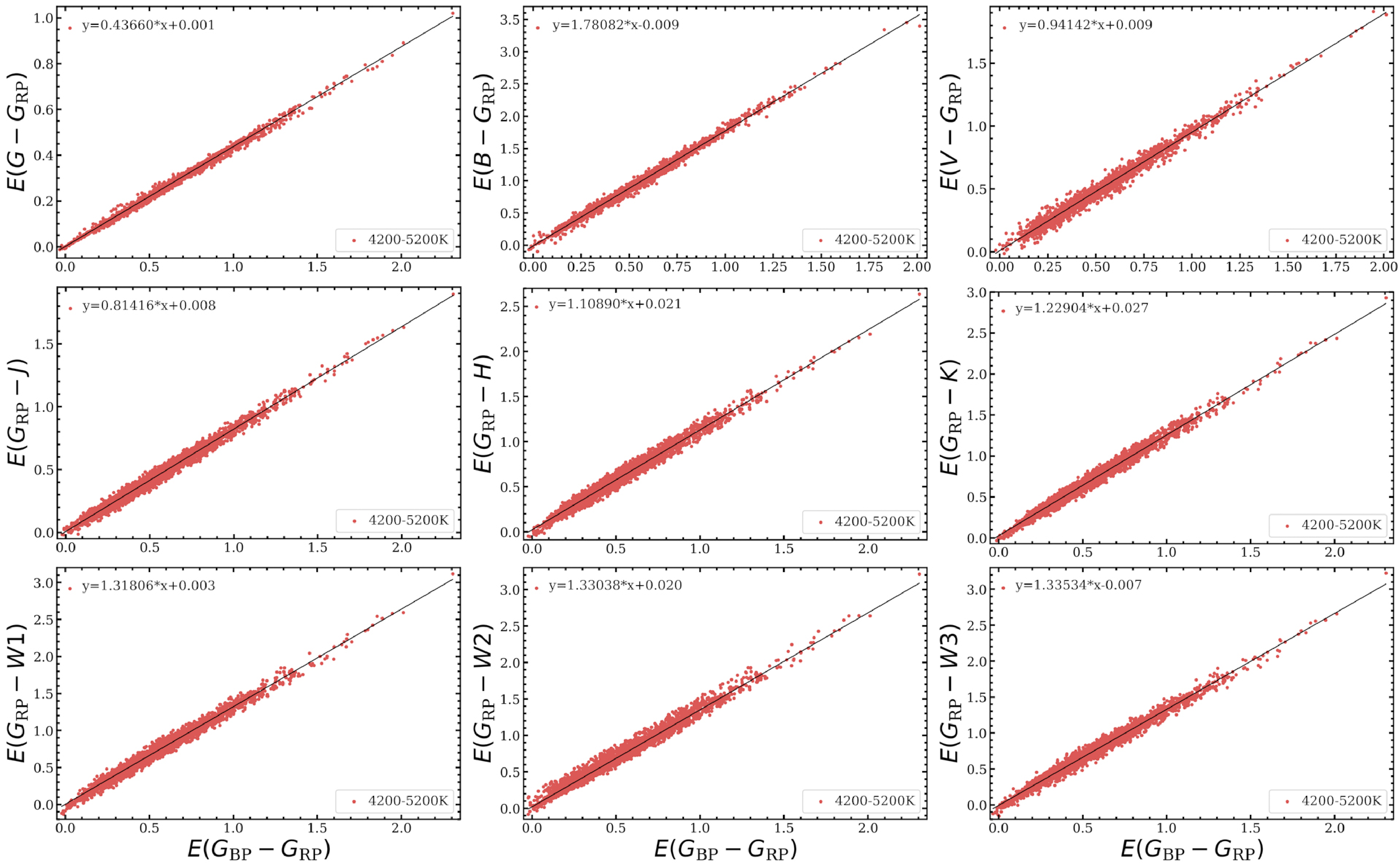}
\caption{CE--CE diagrams $E(\GBP-\GRP)$ vs. $E(\lambda-\GRP)$ of the Taurus giants, where $\lambda$ are $G$ band from Gaia, $B$ and $V$ bands from APASS, $J, H$, and $\Ks$ bands from 2MASS, $W1, W2$, and $W3$ bands from WISE, respectively, from the top left to the bottom right. The black lines are the best linear fit lines, and the results are listed in the upper left corner of each subfigure.}
\label{fig:giant CE}
\end{figure}

\begin{figure}[htbp]
\centering
\includegraphics[scale=0.25]{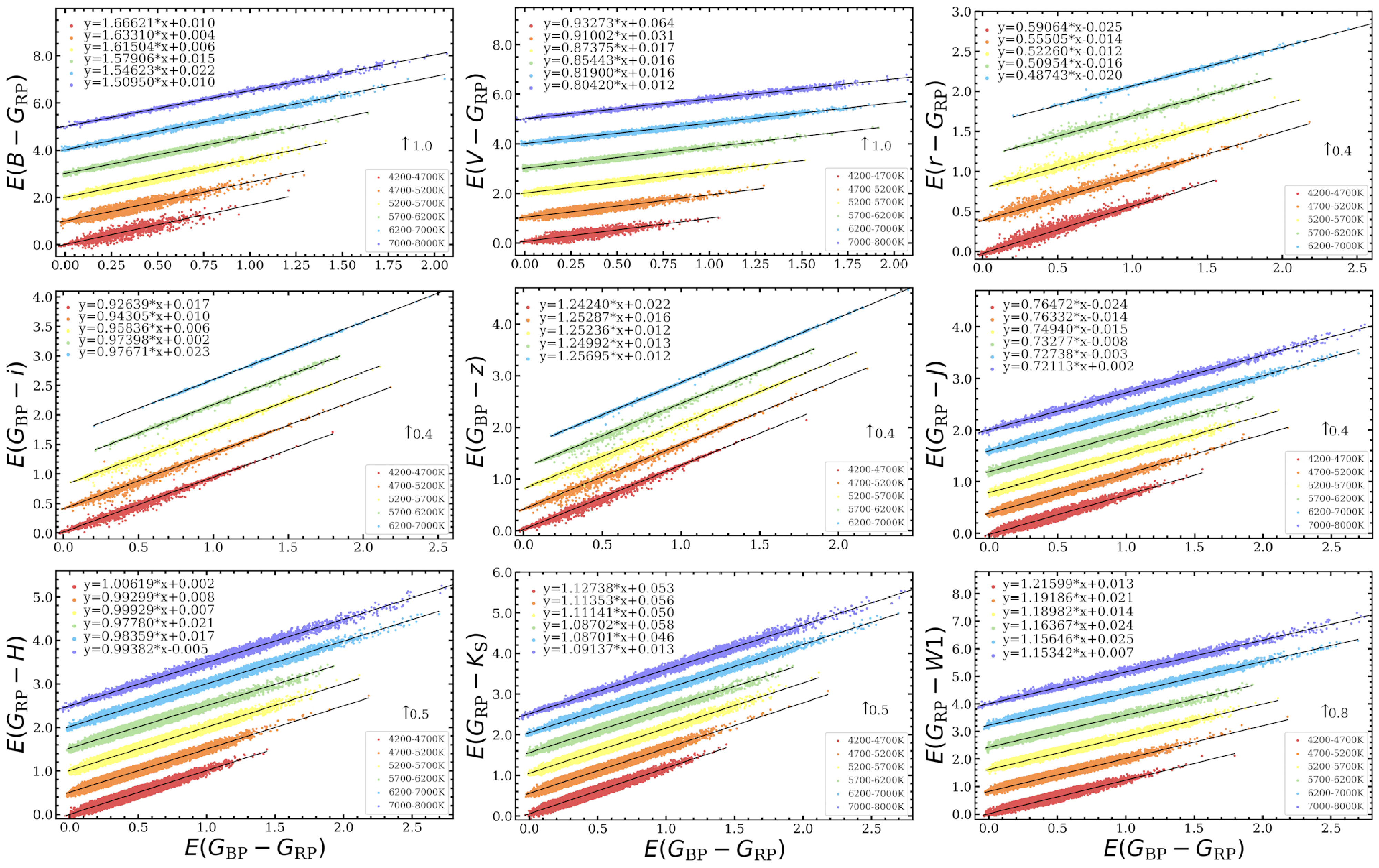}
\caption{CE--CE diagrams $E(\GBP-\GRP)$ vs. $E(\lambda-\GRP)$ or $E(\lambda-\GBP)$ of the Taurus dwarfs with $-0.5\leq{\rm[M/H]}< 0$, where $\lambda$ is $B$ and $V$ bands from APASS, $r, i$, and $z$ bands from PS1, $J, H$ and $\Ks$ bands from 2MASS, $W1$ band from WISE, respectively, from the top left to the bottom right. The black lines are the best linear fit lines, and the results are listed in the upper left corner of each subfigure.}
\label{fig:D2 CE}
\end{figure}

After the curvature correction, we calculate the CERs. Figure~\ref{fig:giant CE} and Figure~\ref{fig:D2 CE} are examples of the determination of the CERs $E(\lambda_1-\GRP)/E(\GBP-\GRP)$ or $E(\lambda_2-\GBP)/E(\GBP-\GRP)$. 
Figure~\ref{fig:giant CE} is for the giant stars, and Figure~\ref{fig:D2 CE} is for the dwarf stars with $-0.5\leq{\rm[M/H]} < 0$, and the CE for some $\Teff$ intervals have been shifted vertically to help distinguish them. 
The measurements of the different bands all exhibit good linearity. The black lines are the best linear fit lines, and the results are shown in the upper left corner of each subfigure. 
The CERs are the slopes of the linear fits and are listed in the third column of Table \ref{tab:AA}.

\subsection{Relative Extinction}\label{sec:Relative Extinction}

The relative extinction $A_\lambda/A_{\GRP}$ can be converted from CERs by 
\begin{equation}\label{equ:AA}
\begin{aligned}
   &A_{{\lambda}_1}/A_{\GRP}=1+k_{{\lambda}_1}(A_{\GBP}/A_{\GRP}-1), {\rm or} \\
   &A_{{\lambda}_2}/A_{\GRP}=A_{\GBP}/A_{\GRP}+k_{{\lambda}_2}(A_{\GBP}/A_{\GRP}-1).
\end{aligned}
\end{equation}
In the conversion, the value of $A_{\GBP}/A_{\GRP}$ is required. 
We adopt the $A_{\GBP}/A_{\GRP}$ value from \cite{2019ApJ...877..116W}. 
However, for different types of stars, the effective wavelengths are slightly different. Therefore, the $A_{\GBP}/A_{\GRP}$ values related to the effective wavelengths of $\GBP$ and $\GRP$ are slightly different. 
We first calculate the static effective wavelength of each band by  
\begin{equation}\label{equ:lambda eff}
\begin{aligned}
   \lambda_{\rm{eff,0}}=\cfrac{\int \lambda F_{\lambda}(\lambda)S(\lambda)d\lambda}{\int  F_{\lambda}(\lambda)S(\lambda)d\lambda}.
\end{aligned}
\end{equation}
The derived $\lambda_{\rm{eff,0}}$ are listed in the second column of Table \ref{tab:AA}. 
Then, for different types of stars, we determined the corresponding $\ABP/\ARP$ values based on the derived $\lambda_{\rm{eff,0}}$ and the extinction law of \cite{2019ApJ...877..116W}. 
Finally, with the obtained $A_{\GBP}/A_{\GRP}$ and CERs, the UV to IR multiband relative extinction $A_\lambda/A_{\GRP}$ is derived by Equation (\ref{equ:AA}). The values of $A_\lambda/A_{\GRP}$ are tabulated in Table~\ref{tab:AA}, and the $A_\lambda/\AV$ and $A_\lambda/E(B-V)$ values are also listed.

\section{Results and Discussion}\label{Result and Discussion}
\subsection{UV--Mid-IR Extinction Curve of the Taurus Region}

\begin{figure}[htbp]
\centering
\includegraphics[scale=0.08]{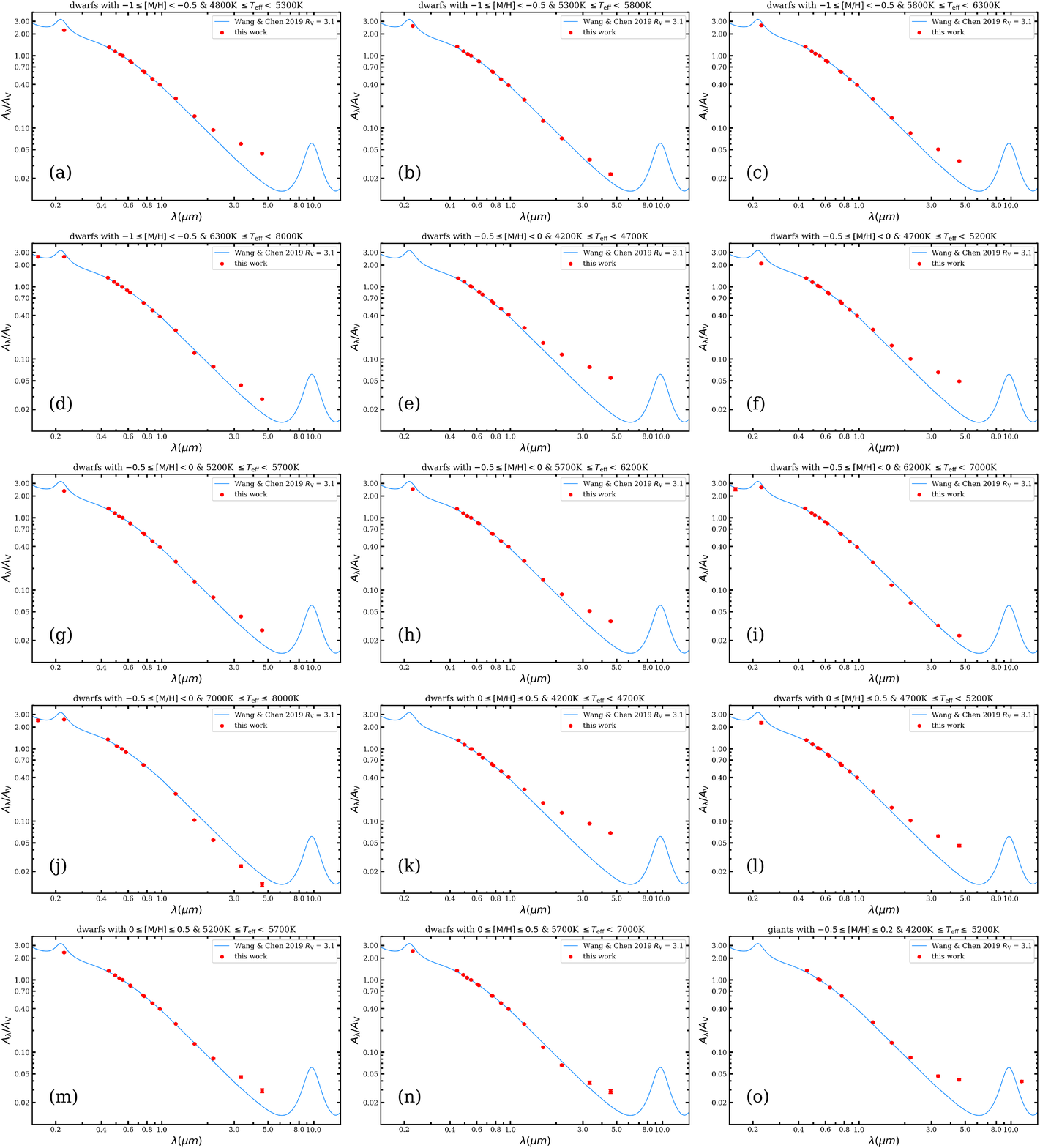}
\caption{UV to mid-IR multiband extinction $A_\lambda$ relative to $\AV$ (red circles with error bars) for different types of stars. The blue lines are the $\RV=3.1$ extinction curve of \cite{2019ApJ...877..116W}.}
\label{fig:ec}
\end{figure}

Based on the different types of dwarfs and giants, we derived the extinction curves of the Taurus region.
Figure~\ref{fig:ec} displays our UV to mid-IR extinction curves, plotted by the red filled circles with error bars. The blue lines are $\RV = 3.1$ extinction law from \cite{2019ApJ...877..116W}. As shown in Figure~\ref{fig:ec}, the extinction curves obtained using different types of stars are consistent in the UV to optical bands.
The UV to optical extinction curves satisfy the $\RV=3.1$ extinction law. In contrast, the IR extinction curves exhibit some variations. The relative extinction values measured with low-temperature dwarfs are slightly larger, as shown in Figures~\ref{fig:ec} (e) and \ref{fig:ec} (k). In the future, we will further analyze the possible variation of the IR extinction law with the James Webb Space Telescope (JWST).

\begin{figure}[htbp]
\centering
\includegraphics[scale=0.15]{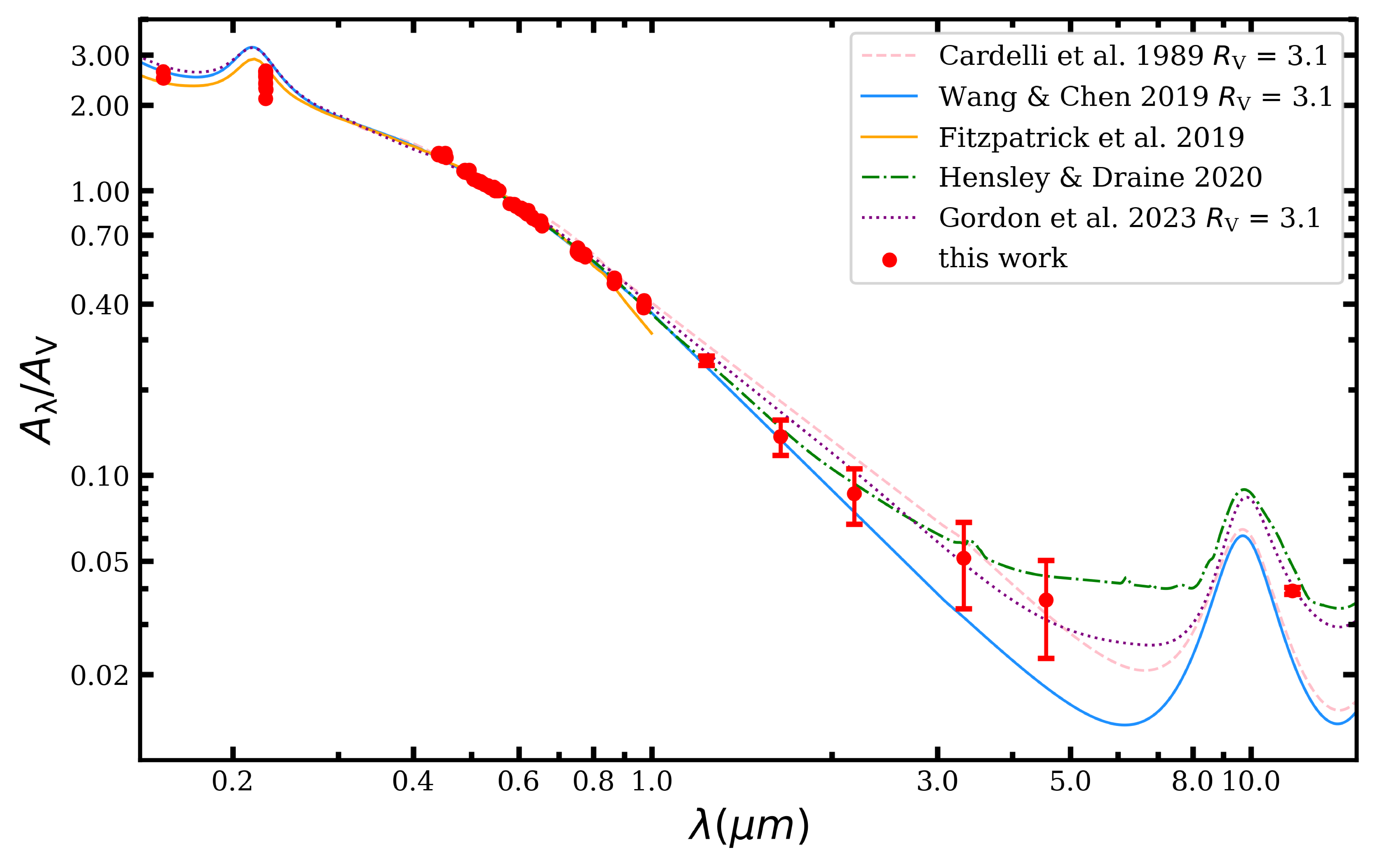}
\caption{The UV to mid-IR average extinction curve of the Taurus molecular cloud (red filled circles with error bars). The color bars in $J, H, \Ks, W1$, and $W2$ bands represent the RMSE of the relative extinction values for the different types of stars. For comparison, the extinction curves of \citet[][pink dashed line]{1989ApJ...345..245C}, \citet[][blue solid line]{2019ApJ...877..116W}, \citet[][orange solid line]{2019ApJ...886..108F}, \citet[][green dot-dashed line]{2020ApJ...895...38H} , and \citet[][purple dotted line]{2023arXiv230401991G} are also shown.}
\label{fig:ec1}
\end{figure}

Combining the relative extinction measured by different types of stars, we obtain the average extinction curve of the Taurus molecular cloud region, as shown in Figure~\ref{fig:ec1}. 
In the optical wavelength of 0.5--1.0$\mum$, since different types of stars have different $\lambda_{\rm eff,0}$, we obtain the extinction curve with a relatively complete wavelength coverage. This is the first attempt to obtain the extinction curve with high wavelength resolution using photometric data.

For the IR bands, the effective wavelengths of the $J, H, \Ks, W1$, and $W2$ bands do not vary significantly for different types of stars as the spectral energy density in IR wavelengths is not sensitive to the effective temperature. We calculate the mean value and root mean square error (RMSE) of the relative extinction values for the different types of stars and show them as the red filled circles and error bars in Figure~\ref{fig:ec1}. 
For the $W3$ band, we only derived the relative extinction $A_{\rm W3}/\AV$ value based on the giant sample. The error bar for this band contains only the internal uncertainty.

In Figure~\ref{fig:ec1}, we also compare our extinction curve with those of \cite{1989ApJ...345..245C}, \cite{2019ApJ...877..116W}, \cite{2019ApJ...886..108F}, \cite{2020ApJ...895...38H}, and \cite{2023arXiv230401991G}, indicated by different colored lines. 
Generally, the reported extinction curves and our extinction values are consistent in the optical range of $\sim0.4-0.6\mum$. 
Our extinction results closely match the extinction curves of \cite{2019ApJ...877..116W} and \cite{2020ApJ...895...38H} in the optical bands ($\sim0.44-1.0\mum$). At $\lambda <0.6\mum$, our extinction results agree well with the curve of \cite{1989ApJ...345..245C}. However, at $\lambda >0.6\mum$, the extinction curve of \cite{1989ApJ...345..245C} is higher than the others. The overestimation of this classical extinction law at long wavelengths was found by \cite{2019ApJ...877..116W} using the Gaia photometry. 

In the IR wavelengths, these extinction curves begin to show significant differences. Compared with the optical band, the extinction in the IR band is smaller, resulting in the error of the IR extinction law usually around 10--20\%. Interestingly, we find the IR extinction curve is steeper when measured using objects with larger extinction. The curves of \cite{2019ApJ...877..116W} and \cite{2020ApJ...895...38H} are both based on objects with extinction $A_V\gtrsim10$ mag, while the curves of \cite{1989ApJ...345..245C} and \cite{2023arXiv230401991G} are based on objects with moderate extinction ($\AV\lesssim4.6$ mag, most $\AV<3$ mag). The average extinction for the Taurus area is $\AV\sim4$ mag, and our IR extinction law lies in the middle of these extinction curves. Nevertheless, the larger error does not allow further optimization of the IR extinction curve.

It is still uncertain whether the IR extinction law is universal or variable. One of the biggest difficulties is how to optimize its accuracy. A sufficiently large extinction is required to obtain a high-precision IR extinction law. However, a large extinction usually leads to a lack of accuracy in the physical parameters of the objects, which can also lead to a large error in the extinction law. JWST will provide more opportunities to optimize the IR extinction law.

\subsection{Variations of Taurus Extinction Law}\label{sec:Variations of Taurus Extinction}

We explore the possible variation of extinction law in the Taurus molecular cloud region.
We divide the whole Taurus region into subregions according to the galactic longitude $l$ and the galactic latitude $b$. Each subregion covers $l\times b = 2^\circ \times 1^\circ$. We adopt dwarf stars with $-0.5 \leq {\rm[M/H]} < 0$ and 4700\,K $\leq \Teff < 5200$\,K to fit the CE--CE plots and obtain the CER $E(\GBP-\Ks)/E(\GBP-\GRP)$ for each subregion. 

We use a similar approach to \cite{2016ApJ...821...78S}
to establish the $\RV$--CER $E(\GBP-\Ks)/E(\GBP-\GRP)$ conversion equation based on the $\RV$-dependent extinction curves \citep{2019ApJ...877..116W,2023ApJ...946...43W}. The equation is
\begin{equation}\label{equ:transRV}
\begin{aligned}
   \RV=&2.517\times[E(\GBP-\Ks)/E(\GBP-\GRP)]^2 \\
   &-6.85\times E(\GBP-\Ks)/E(\GBP-\GRP)\\
   &+6.107.
\end{aligned}
\end{equation}
According to this equation, we convert our determined $E(\GBP-\Ks)/E(\GBP-\GRP)$ into $\RV$ value and analyze the variations of $\RV$ in the Taurus molecular cloud.

\begin{figure}[htbp]
\centering
\includegraphics[scale=0.4]{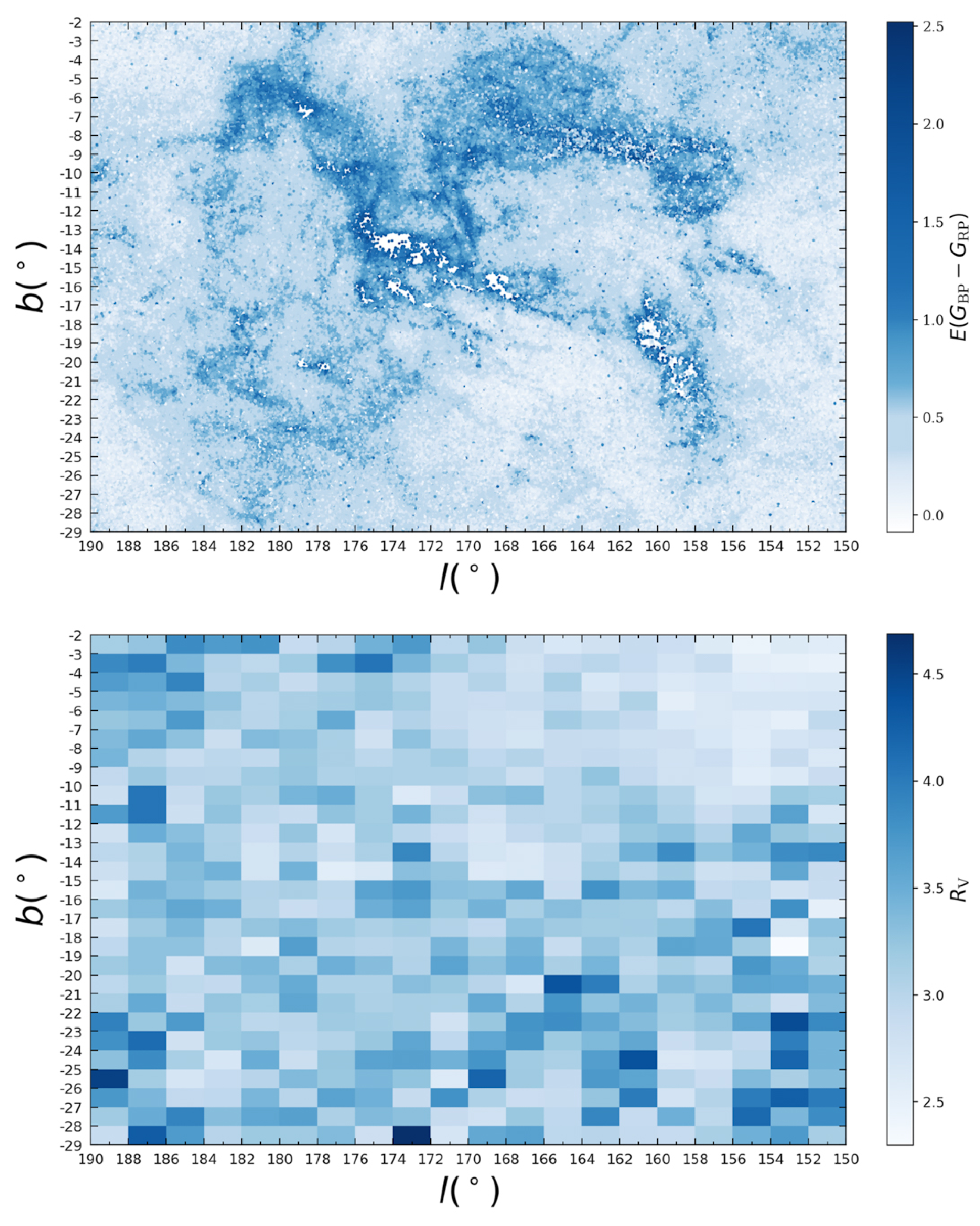}
\caption{Top: Extinction map of the Taurus molecular cloud region colored by the CE $E(\GBP-\GRP)$.
Bottom: Variation of extinction law in the Taurus molecular cloud region characterized by $\RV$. The entire Taurus region are divided into subregions with a grid of $l\times b$=$2^\circ \times 1^\circ$. 
There is no correlation between the regions with large deviations in $\RV$ and the amount of the CE $E(\GBP-\GRP)$ or any structure of the Taurus cloud.
}
\label{fig:local area CE}
\end{figure}

The top panel of Figure~\ref{fig:local area CE} displays an extinction map of the Taurus molecular cloud colored by the CE $E(\GBP-\GRP)$. The larger the $E(\GBP-\GRP)$, i.e., the higher the extinction, the darker the color. 
In the core areas of the Taurus cloud, e.g., $l=174^\circ$, $b=-14^\circ$, there is no CE value due to lack of data. 
The bottom panel of Figure~\ref{fig:local area CE} shows the variation of $\RV$. 
As shown in Figure~\ref{fig:local area CE}, there is no correlation between the regions with large deviations in $\RV$ and the amount of the CE $E(\GBP-\GRP)$ or any structure of the Taurus cloud. This suggests that the optical extinction law of the Taurus region is almost invariant.

\begin{figure}[htbp]
\centering
\includegraphics[scale=0.2]{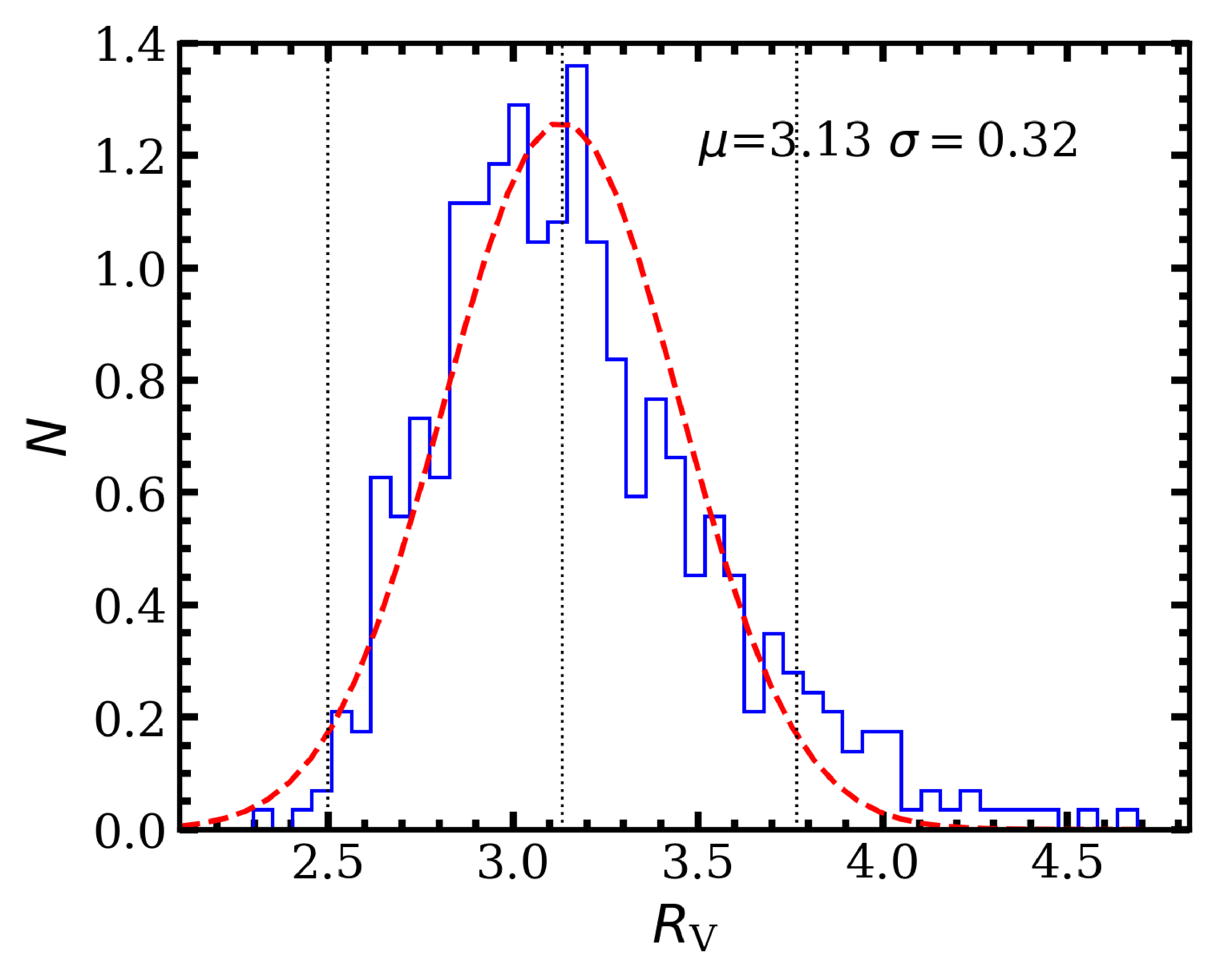}
\caption{Distribution of $\RV$ for each subregion of the Taurus molecular cloud. This distribution is fitted by a Gaussian function with a mean value of $\mu=3.13$ and a width of $\sigma$ = 0.32. The dotted lines show the 16th, 50th, and 84th percentiles of the $\RV$ distribution.}
\label{fig:gauss}
\end{figure}

Figure~\ref{fig:gauss} is the histogram of $\RV$, and its distribution can be well fitted by a Gaussian function with a mean value of $\mu=3.13$ and a width of $\sigma=0.32$ (the red dashed line). The percentage of subregions with $\RV$ deviations $>3\sigma$ is only 0.9\%. The scatter of our $\RV$ is larger than that of \cite{2016ApJ...821...78S} for the Galactic disk ($\sigma=0.18$). The smaller dispersion of their $\RV$ is due to the fact that they used the APOGEE data, which is more accurate, and the extinction of the object is larger.

\subsection{Comparison with Gaia CE}\label{sec:comp Gaia} 

\begin{figure}[htbp]
\centering
\includegraphics[scale=0.3]{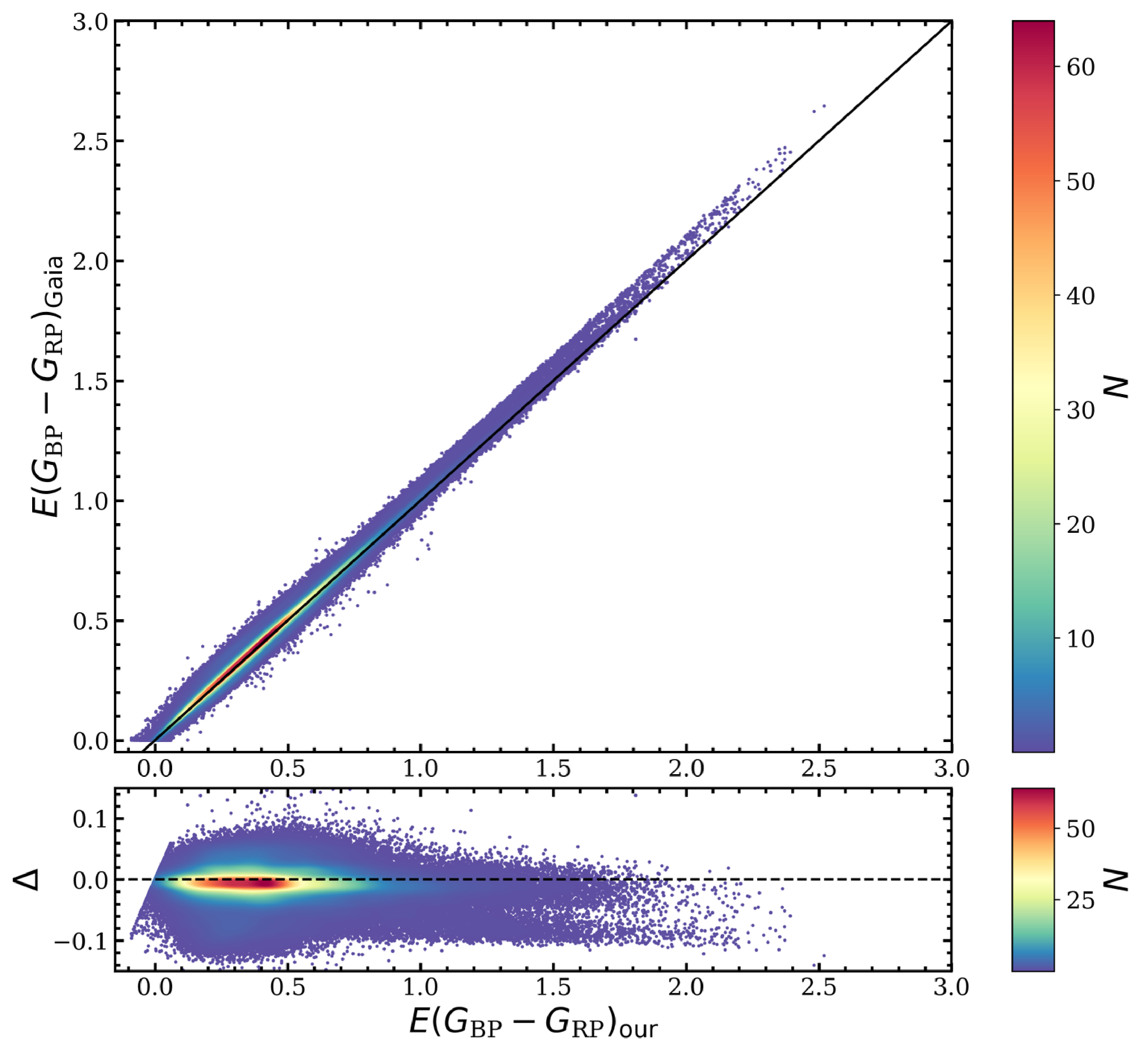}
\caption{Comparison of our $E(\GBP-\GRP)$ with Gaia $E(\GBP-\GRP)$, colored by the number density of stars. The x-axis is the $E(\GBP-\GRP)$ derived in this work, the y-axis is the $E(\GBP-\GRP)$ from Gaia DR3, and the black solid line is $y = x$.  The bottom panel is the distribution of CE residuals $\Delta$ = $E(\GBP-\GRP)_{\rm our}-E(\GBP-\GRP)_{\rm Gaia}$, and the black dashed line is $y = 0$.}
\label{fig:comp Gaia}
\end{figure}

Gaia DR3 catalog also provides CE $E(\GBP-\GRP)$. We compared the Gaia DR3 $E(\GBP-\GRP)$ with our determined $E(\GBP-\GRP)$. 
Figure~\ref{fig:comp Gaia} shows the comparison, the colors indicating the number density of the sample. 
In the top panel of Figure~\ref{fig:comp Gaia}, the x-axis is our CE, the y-axis is the CE from Gaia, and the solid black line is the one-to-one line $y = x$. 
The bottom panel of Figure~\ref{fig:comp Gaia} displays the distribution of the CE residuals $\Delta$ = $E(\GBP-\GRP)_{\rm our}-E(\GBP-\GRP)_{\rm Gaia}$, where the black dashed line is $y = 0$. 

At $E(\GBP-\GRP)<1.3$ mag, our CEs are in good agreement with Gaia CEs. At $E(\GBP-\GRP)>1.5$ mag, the Gaia CEs are overestimated. 
In Figure~\ref{fig:comp Gaia}, it is clear that the overestimation of Gaia CE becomes more significant as the CE increases. 
However, there are very few stars with large deviations. Only 0.5\% of stars have $\Delta>0.1$ mag, and 92\% of stars have $E(\GBP-\GRP)< 1.3$ mag with $\Delta<0.05$ mag.

As mentioned in Section~\ref{sec:Data}, there is a degeneracy between the Gaia $\Teff$ estimation and extinction. The higher the extinction, the more significant the overestimation of Gaia $\Teff$.
This means that a CE overestimation will lead to an overestimation of $\Teff$. 
From Figure~\ref{fig:comp Gaia}, we confirm that there is an overestimation of Gaia CE and use a polynomial fit to obtain the CE overestimation $\Delta$ (CE residuals) of each star. 
The CE overestimation is equivalent to the intrinsic color deviation, i.e., the larger the CE, the smaller the intrinsic color. 
In Section~\ref{sec:intrinsic color}, we have established the $\Teff$--intrinsic color relations. 
Using the determined $\Teff$--$(\GBP-\GRP)_0$ relation, we obtain the $\Teff$ deviation from the intrinsic color deviation. After that, we correct the CE overestimation and re-estimate the $\Teff$. 
The re-estimated Gaia $\Teff$ is more consistent with the APOGEE $\Teff$, with the mean differences for giants and dwarfs reduced to $\sim91 \K$ and $\sim13 \K$, respectively.

We further analyze the effects of the CE overestimation on the determination of the Taurus extinction law.
Based on the newly estimated $\Teff$, we reclassify the stars using the same procedure as in Section~\ref{sec:Color Excess Ratio} and determine the multiband CERs for the different types of stars. 
Compared to the CERs derived in Section~\ref{sec:Color Excess Ratio}, the new CERs show good agreement, with only 4\% of stars having CER differences $> 2\sigma$. 
Similar to the method described in Section~\ref{sec:Variations of Taurus Extinction} and Figure~\ref{fig:gauss}, we calculate the $\RV$ of each subregion of the Taurus cloud and analyze the $\RV$ distribution. The new average $\RV$ fitted by a Gaussian function increases slightly, from 3.13 to 3.14.
Therefore, we conclude that the bias in $\Teff$ caused by the overestimation of the Gaia CE has little effect on the determined extinction law.

\section{Conclusion}\label{Summary}

With dwarfs and giants selected by the stellar parameters of Gaia DR3, we have investigated the extinction curves of UV to mid-IR in the Taurus molecular cloud.
The multiband photometric data are collected from GALEX, Gaia, APASS, PS1, 2MASS, and WISE surveys.
The main results of this work are as follows:
\begin{enumerate}
\item We established the mid-IR bands $\Teff$--intrinsic color relations and determined the intrinsic color indices for different types of stars based on Gaia stellar parameters. We determined multiband CEs, including the GALEX (FUV and NUV), Gaia ($G, \GBP, \GRP$), APASS ($B, V$), PS1 ($g, r, i, z, y$), 2MASS ($J, H, \Ks$), WISE ($W1, W2, W3$) bands.
\item In the determination of CERs by using the CE method, we considered the curvature of CERs caused by the assumption of a static wavelength for stars at each filter in fitting the  CE--CE diagrams. For different types of stars, we first calibrated the curvature of CERs. After curvature corrections, the final CERs are determined and listed in Table~\ref{tab:AA}. 
\item After converting the CERs to relative extinction, we obtained the average extinction curve of the Taurus molecular cloud. In UV to near-IR bands ($\sim0.15-2.5\mum$), our extinction curve is well consistent with the $\RV=3.1$ average extinction law of the Milky Way. The extinction law in the longer IR bands are still controversial and need further investigation. 
\item Using dwarfs with $-0.5 \leq {\rm [M/H]} < 0$ and 4700\,K $\leq \Teff < 5200$\,K, we investigated the possible regional variation of extinction law in the Taurus molecular cloud region. We derive a mean $\RV=3.13\pm0.32$ for the Taurus subregions, with only 0.9\% of subregions having deviations $>3\sigma$, and 3\% of the subregions having $\RV>4$. We conclude the optical extinction law of the Taurus region is almost invariant.
\item Compared with Gaia CE, we found that at $E(\GBP-\GRP)>1.5$ mag, the Gaia CEs are overestimated. The overestimation in Gaia CE affects the determination of Gaia stellar parameter values, such as $\Teff$. 
We used our established $\Teff$--intrinsic color relations to analyze the effects of the CE overestimation on the determination of $\Teff$. 
We found that the higher the extinction, the more significant the overestimation of $\Teff$.
After correcting the CE overestimation and re-estimating $\Teff$, we also analyze the possible effects of the CE overestimation on the extinction law. It was confirmed that the overestimated $\Teff$ hardly affects our extinction results. 
\end{enumerate}

\section*{Acknowledgements}

We thank Drs. He Zhao and Mingxu Sun for very helpful discussions, and the anonymous referee for very useful comments/suggestions. This work is supported by the National Natural Science Foundation of China (NSFC) through the projects 12003046, 12173047, 11903045, and 12133002. This work is also supported by the National Key Research and Development Program of China, grant 2019YFA0405504 and the science research grants from the China Manned Space Project with No. CMS-CSST-220221-A09. S.W. and X.C. acknowledge support from the Youth Innovation Promotion Association of the CAS (grant No. 2023065 and 2022055). Q.J. thank the support from the Sichuan Youth Science and Technology Innovation Research Team (21CXTD0038).

\bibliography{Taurus_reference}{}
\bibliographystyle{aasjournal}

%1

%4X3  2

%3

%4

%3X3 giant CE  5

%3X3 D2 CE  6

%7

%8

%9

%10

%11

\clearpage

\begin{center} 
\begin{table}
\caption{Different Samples Used in This Work}
\label{tab:sample}
\begin{tabular}{lccc}
\hline\hline\noalign{\smallskip}
Name         & Notes                                                                 & Sample size    & Section \\
\hline 
Dwarf sample & $\log g \geq 4$ \& 4200\,K $\leq \Teff \leq 8000$\,K \& $-1 \leq {\rm[M/H]} \leq 0.5$        & 504,988      & \multirow{2}{*}{\ref{sec:The Giant and Dwarf Samples}}      \\
Giant sample & $1 \leq \log g \leq 3.3$ \& 4200\,K $\leq \Teff \leq 5200$\,K \& $-0.5 \leq {\rm[M/H]} \leq 0.2$ & 4,757          &        \\
\hline 
\multirow{15}{*}{Different types of stars} & $\log g \geq 4$ \& $-1 \leq {\rm[M/H]} < -0.5$ \& 4800\,K $\leq \Teff < 5300$\,K & 41,719  & \multirow{15}{*}{\ref{sec:Color Excess Ratio}} \\
             & $\log g \geq 4$ \& $-1 \leq {\rm[M/H]} < -0.5$ \& 5300\,K $\leq \Teff < 5800$\,K     & 34,092         &         \\
             & $\log g \geq 4$ \& $-1 \leq {\rm[M/H]} < -0.5$ \& 5800\,K $\leq \Teff < 6300$\,K     & 26,421         &         \\
             & $\log g \geq 4$ \& $-1 \leq {\rm[M/H]} < -0.5$ \& 6300\,K $\leq \Teff \leq 8000$\,K     & 12,676         &         \\
             & $\log g \geq 4$ \& $-0.5 \leq {\rm[M/H]} < 0$ \& 4200\,K $\leq \Teff < 4700$\,K      & 74,533         &         \\
             & $\log g \geq 4$ \& $-0.5 \leq {\rm[M/H]} < 0$ \& 4700\,K $\leq \Teff < 5200$\,K      & 55,288         &         \\
             & $\log g \geq 4$ \& $-0.5 \leq {\rm[M/H]} < 0$ \& 5200\,K $\leq \Teff < 5700$\,K      & 53,995         &         \\
             & $\log g \geq 4$ \& $-0.5 \leq {\rm[M/H]} < 0$ \& 5700\,K $\leq \Teff < 6200$\,K      & 41,609         &         \\
             & $\log g \geq 4$ \& $-0.5 \leq {\rm[M/H]} < 0$ \& 6200\,K $\leq \Teff < 7000$\,K      & 22,548         &         \\
             & $\log g \geq 4$ \& $-0.5 \leq {\rm[M/H]} < 0$ \& 7000\,K $\leq \Teff \leq 8000$\,K   & 5,777          &         \\
             & $\log g \geq 4$ \& $0 \leq {\rm[M/H]} \leq 0.5$ \& 4200\,K $\leq \Teff < 4700$\,K       & 75,874         &         \\
             & $\log g \geq 4$ \& $0 \leq {\rm[M/H]} \leq 0.5$ \& 4700\,K $\leq \Teff < 5200$\,K       & 8,643          &         \\
             & $\log g \geq 4$ \& $0 \leq {\rm[M/H]} \leq 0.5$ \& 5200\,K $\leq \Teff < 5700$\,K       & 3,472          &         \\
             & $\log g \geq 4$ \& $0 \leq {\rm[M/H]} \leq 0.5$ \& 5700\,K $\leq \Teff < 7000$\,K       & 1,405          &         \\
             & $1 \leq \log g \leq 3.3$ \& 4200\,K $\leq \Teff \leq 5200$\,K \& $-0.5 \leq {\rm[M/H]} \leq 0.2$   & 4,757          &    \\ 
\noalign{\smallskip}\hline
\end{tabular}
\end{table}
\end{center}

\clearpage

\begin{center} 
\begin{longtable}{lccccc}
\caption{Multiband Color Excess Ratios and Extinction Coefficients}
\label{tab:AA} \\
\hline\hline\noalign{\smallskip}
Band & $\lambda_{\rm eff, 0}$ ($\mum$) & $\frac{E(\lambda-\GRP)}{E(\GBP-\GRP)}$ & $A_\lambda/\ARP$  & $A_\lambda/\AV$ & $A_\lambda/E(B-V)$ \\ 
\endfirsthead
\multicolumn{6}{c}
{{\bfseries \tablename\ \thetable{} -- continued}} \\
\hline\hline\noalign{\smallskip}
Band & $\lambda_{\rm eff, 0}$ ($\mum$) & $\frac{E(\lambda-\GRP)}{E(\GBP-\GRP)}$ & $A_\lambda/\ARP$  & $A_\lambda/\AV$ & $A_\lambda/E(B-V)$ \\ 
\hline 
\endhead			
\endfoot
\hline
\endlastfoot

\noalign{\smallskip}\hline\noalign{\smallskip}
\multicolumn{6}{c}{dwarfs with $-1 \leq {\rm[M/H]} < -0.5$ \& 4800\,K $\leq \Teff < 5300$\,K} \\
\hline 
NUV     & 0.2270 & $3.725\pm0.086$  & $3.815\pm0.065$ & $2.267\pm0.039$ & 7$.089\pm0.135$ \\
$\GBP$  & 0.5317 & 1            & $1.756\pm0.000$ & $1.043\pm0.001$ & $3.263\pm0.028$ \\
$\GRP$  & 0.7680 & ...           & 1           & $0.594\pm0.001$ & $1.859\pm0.016$ \\
$G$     & 0.6304 & $0.486\pm0.000$  & $1.367\pm0.000$ & $0.812\pm0.0$01 & $2.541\pm0.022$ \\
$B$     & 0.4483 & $1.616\pm0.005$  & $2.221\pm0.00$4 & $1.320\pm0.003$ & $4.127\pm0.036$ \\
$V$     & 0.5516 & $0.903\pm0.003 $ & $1.683\pm0.00$2 & 1           & $3.127\pm0.027$ \\
$g$     & 0.4923 & $1.258\pm0.000$  & $1.951\pm0.000$ & $1.159\pm0.002$ & $3.625\pm0.031$ \\
$r$     & 0.6205 & $0.546\pm0.000$  & $1.413\pm0.000$ & $0.840\pm0.001$ & $2.626\pm0.022$ \\
$i$     & 0.7520 & $0.054\pm0.00$0  & $1.041\pm0.00$0 & $0.619\pm0.001$ & $1.935\pm0.017$ \\
$z$     & 0.8664 & $-0.253\pm0.000$ & $0.809\pm0.000$ & $0.481\pm0.001$ & $1.503\pm0.013$ \\
$y$     & 0.9702 & $-0.441\pm0.000$ & $0.667\pm0.00$0 & $0.396\pm0.001$ & $1.239\pm0.011$ \\
$J$     & 1.2339 & $-0.749\pm0.001$ & $0.434\pm0.001$ & $0.258\pm0.001$ & $0.806\pm0.007$ \\
$H$     & 1.6390 & $-0.998\pm0.001$ & $0.246\pm0.001$ & $0.146\pm0.001$ & $0.456\pm0.004$ \\
$\Ks$   & 2.1763 & $-1.114\pm0.001$ & $0.158\pm0.00$1 & $0.094\pm0.001$ & $0.294\pm0.003$ \\
$W1$    & 3.3168 & $-1.189\pm0.001$ & $0.102\pm0.001$ & $0.060\pm0.001$ & $0.189\pm0.003$ \\
$W2$    & 4.5585 & $-1.225\pm0.002$ & $0.074\pm0.001$ & $0.044\pm0.001$ & $0.138\pm0.003$ \\
\noalign{\smallskip}\hline
\multicolumn{6}{c}{dwarfs with $-1 \leq {\rm[M/H]} < -0.5$ \& 5300\,K $\leq \Teff < 5800$\,K} \\
\hline 
NUV     & 0.2270 & $4.256\pm0.018$  & $4.355\pm0.014$ & $2.590\pm0.009$ & $7.517\pm0.038$ \\
$\GBP$  & 0.5222 & 1            & $1.788\pm0.000$ & $1.064\pm0.001$ & $3.087\pm0.012$ \\
$\GRP$  & 0.7654 & ...           & 1           & $0.595\pm0.000$ & $1.726\pm0.007$ \\
$G$     & 0.6175 & $0.521\pm0.000$  & $1.411\pm0.000$ & $0.839\pm0.001$ & $2.436\pm0.009$ \\
$B$     & 0.4447 & $1.599\pm0.002$  & $2.261\pm0.002$ & $1.345\pm0.002$ & $3.902\pm0.015$ \\
$V$     & 0.5502 & $0.864\pm0.002$  & $1.681\pm0.001$ & 1           & $2.902\pm0.012$ \\
$g$     & 0.4895 & $1.211\pm0.000$  & $1.955\pm0.000$ & $1.162\pm0.001$ & $3.374\pm0.013$ \\
$r$     & 0.6199 & $0.518\pm0.000$  & $1.409\pm0.000$ & $0.838\pm0.001$ & $2.432\pm0.009$ \\
$i$     & 0.7515 & $0.042\pm0.000$  & $1.033\pm0.000$ & $0.614\pm0.001$ & $1.783\pm0.007$ \\
$z$     & 0.8660 & $-0.254\pm0.000$ & $0.800\pm0.000$ & $0.476\pm0.000$ & $1.380\pm0.005$ \\
$y$     & 0.9697 & $-0.434\pm0.000$ & $0.658\pm0.000$ & $0.391\pm0.000$ & $1.135\pm0.004$ \\
$J$     & 1.2328 & $-0.743\pm0.001$ & $0.414\pm0.001$ & $0.246\pm0.000$ & $0.715\pm0.003$ \\
$H$     & 1.6385 & $-1.002\pm0.001$ & $0.210\pm0.001$ & $0.125\pm0.000$ & $0.363\pm0.002$ \\
$\Ks$   & 2.1762 & $-1.115\pm0.001$ & $0.121\pm0.001$ & $0.072\pm0.001$ & $0.208\pm0.002$ \\
$W1$    & 3.3168 & $-1.191\pm0.001$ & $0.061\pm0.001$ & $0.036\pm0.001$ & $0.106\pm0.002$ \\
$W2$    & 4.5609 & $-1.219\pm0.001$ & $0.039\pm0.001$ & $0.023\pm0.001$ & $0.067\pm0.002$ \\
\noalign{\smallskip}\hline
\multicolumn{6}{c}{dwarfs with $-1 \leq {\rm[M/H]} < -0.5$ \& 5800\,K $\leq \Teff < 6300$\,K}  \\
\hline 
NUV     & 0.2270 & $4.249\pm0.015$  & $4.413\pm0.012$ & $2.631\pm0.0$08 & $7.731\pm0.041$ \\
$\GBP$  & 0.5146 & 1            & $1.803\pm0.000$ & $1.075\pm0.001$ & $3.159\pm0.014$ \\
$\GRP$  & 0.7626 & ...           & 1           & $0.596\pm0.001$ & $1.752\pm0.008$ \\
$G$     & 0.6043 & $0.551\pm0.002$  & $1.442\pm0.002$ & $0.860\pm0.001$ & $2.527\pm0.012$ \\
$B$     & 0.4424 & $1.554\pm0.003$  & $2.248\pm0.002$ & $1.340\pm0.002$ & $3.938\pm0.018$ \\
$V$     & 0.5491 & $0.843\pm0.002$  & $1.677\pm0.002$ & 1           & $2.938\pm0.013$ \\
$g$     & 0.4874 & $1.177\pm0.000$  & $1.945\pm0.000$ & $1.160\pm0.001$ & $3.408\pm0.015$ \\
$r$     & 0.6188 & $0.497\pm0.001$  & $1.399\pm0.00$1 & $0.834\pm0.001$ & $2.451\pm0.011$ \\
$i$     & 0.7509 & $0.029\pm0.001$  & $1.023\pm0.001$ & $0.610\pm0.001$ & $1.792\pm0.008$ \\
$z$     & 0.8658 & $-0.248\pm0.001$ & $0.801\pm0.001$ & $0.477\pm0.001$ & $1.403\pm0.006$ \\
$y$     & 0.9696 & $-0.420\pm0.001$ & $0.663\pm0.000$ & $0.395\pm0.000$ & $1.161\pm0.005$ \\
$J$     & 1.2321 & $-0.719\pm0.001$ & $0.422\pm0.001$ & $0.252\pm0.000$ & $0.740\pm0.003$ \\
$H$     & 1.6379 & $-0.956\pm0.001$ & $0.232\pm0.00$1 & $0.139\pm0.001$ & $0.407\pm0.002$ \\
$\Ks$   & 2.1761 & $-1.068\pm0.001$ & $0.142\pm0.001$ & $0.085\pm0.001$ & $0.249\pm0.002$ \\
$W1$    & 3.3167 & $-1.139\pm0.00$1 & $0.085\pm0.00$1 & $0.051\pm0.001$ & $0.149\pm0.002$ \\
$W2$    & 4.5630 & $-1.172\pm0.001$ & $0.059\pm0.001$ & $0.035\pm0.001$ & $0.103\pm0.002$ \\
\noalign{\smallskip}\hline
\multicolumn{6}{c}{dwarfs with $-1 \leq {\rm[M/H]} < -0.5$ \& 6300\,K $\leq \Teff \leq 8000$\,K} \\
\hline
FUV     & 0.1531 & $4.099\pm0.132$  & $4.389\pm0.109$ & $2.626\pm0.065$ & $7.727\pm0.194$ \\
NUV     & 0.2268 & $4.091\pm0.011$  & $4.382\pm0.009$ & $2.621\pm0.006$ & $7.714\pm0.029$ \\
$\GBP$  & 0.5077 & 1            & $1.827\pm0.000$ & $1.093\pm0.001$ & $3.216\pm0.010$ \\
$\GRP$  & 0.7585 & ...           & 1           & $0.598\pm0.000$ & $1.760\pm0.005$ \\
$G$     & 0.5889 & $0.607\pm0.000$  & $1.502\pm0.000$ & $0.898\pm0.001$ & $2.643\pm0.008$ \\
$B$     & 0.4407 & $1.499\pm0.002$  & $2.240\pm0.001$  & $1.340\pm0.001$ & $3.943\pm0.012$ \\
$V$     & 0.5477 & $0.812\pm0.001$  & $1.672\pm0.001$ & 1           & $2.943\pm0.009$ \\
$g$     & 0.4850 & $1.166\pm0.002$  & $1.964\pm0.001$ & $1.175\pm0.001$ & $3.457\pm0.011$ \\
$r$     & 0.6174 & $0.484\pm0.003$  & $1.400\pm0.002$ & $0.837\pm0.00$2 & $2.464\pm0.009$ \\
$z$     & 0.8655 & $-0.253\pm0.003$ & $0.791\pm0.002$ & $0.473\pm0.001$ & $1.392\pm0.006$ \\
$y$     & 0.9691 & $-0.426\pm0.002$ & $0.648\pm0.001$ & $0.388\pm0.001$ & $1.141\pm0.004$ \\
$J$     & 1.2302 & $-0.703\pm0.001$ & $0.419\pm0.000$ & $0.250\pm0.000$ & $0.737\pm0.002$ \\
$H$     & 1.6373 & $-0.964\pm0.001$ & $0.203\pm0.001$ & $0.121\pm0.000$ & $0.357\pm0.002$ \\
$\Ks$   & 2.1759 & $-1.050\pm0.001$ & $0.132\pm0.001$ & $0.079\pm0.000$ & $0.232\pm0.002$ \\
$W1$    & 3.3167 & $-1.121\pm0.001$ & $0.073\pm0.001$ & $0.044\pm0.001$ & $0.128\pm0.002$ \\
$W2$    & 4.5645 & $-1.154\pm0.001$ & $0.046\pm0.001$ & $0.028\pm0.001$ & $0.081\pm0.002$ \\
\noalign{\smallskip}\hline
\multicolumn{6}{c}{dwarfs with $-0.5 \leq {\rm[M/H]} < 0$ \& 4200\,K $\leq \Teff < 4700$\,K}  \\
\hline 
$\GBP$ & 0.5453 & 1            & $1.716\pm0.000$ & $1.029\pm0.004$ & $3.267\pm0.086$ \\
$\GRP$ & 0.7736 & ...           & 1           & $0.600\pm0.002$ & $1.904\pm0.050$ \\
$G$    & 0.6532 & $0.430\pm0.000$  & $1.308\pm0.000$ & $0.784\pm0.003$ & $2.489\pm0.065$ \\
$B$    & 0.4527 & $1.666\pm0.017$  & $2.193\pm0.012$ & $1.315\pm0.009$ & $4.175\pm0.112$ \\
$V$    & 0.5556 & $0.933\pm0.008$  & $1.668\pm0.006$ & 1           & $3.175\pm0.084$ \\
$g$    & 0.4958 & $1.354\pm0.001$  & $1.969\pm0.000$ & $1.181\pm0.004$ & $3.749\pm0.099$ \\
$r$    & 0.6220 & $0.591\pm0.000$  & $1.423\pm0.000$ & $0.853\pm0.003$ & $2.709\pm0.071$ \\
$i$    & 0.7531 & $0.074\pm0.000$  & $1.053\pm0.000$ & $0.631\pm0.002$ & $2.004\pm0.053$ \\
$z$    & 0.8671 & $-0.242\pm0.000$ & $0.826\pm0.000$ & $0.495\pm0.002$ & $1.573\pm0.041$ \\
$y$    & 0.9708 & $-0.437\pm0.000$ & $0.687\pm0.000$ & $0.412\pm0.002$ & $1.308\pm0.034$ \\
$J$    & 1.2348 & $-0.765\pm0.001$ & $0.452\pm0.001$ & $0.271\pm0.001$ & $0.861\pm0.023$ \\
$H$    & 1.6399 & $-1.006\pm0.001$ & $0.279\pm0.001$ & $0.168\pm0.001$ & $0.532\pm0.014$ \\
$\Ks$  & 2.1761 & $-1.127\pm0.001$ & $0.193\pm0.001$ & $0.115\pm0.001$ & $0.367\pm0.010$ \\
$W1$   & 3.3175 & $-1.216\pm0.001$ & $0.129\pm0.001$ & $0.077\pm0.001$ & $0.246\pm0.007$ \\
$W2$   & 4.5547 & $-1.269\pm0.002$ & $0.091\pm0.001$ & $0.055\pm0.001$ & $0.174\pm0.005$ \\
\noalign{\smallskip}\hline
\multicolumn{6}{c}{dwarfs with $-0.5 \leq {\rm[M/H]} < 0$ \& 4700\,K $\leq \Teff < 5200$\,K} \\
\hline 
NUV     & 0.2270 & $3.406\pm0.085$  & $3.543\pm0.064$ & $2.110\pm0.038$ & $6.562\pm0.130$ \\
$\GBP$  & 0.5327 & 1            & $1.747\pm0.000$ & $1.040\pm0.001$ & $3.235\pm0.027$ \\
$\GRP$  & 0.7675 & ...           & 1           & $0.595\pm0.001$ & $1.852\pm0.015$ \\
$G$     & 0.6310 & $0.475\pm0.000$  & $1.355\pm0.000$ & $0.807\pm0.001$ & $2.509\pm0.021$ \\
$B$     & 0.4487 & $1.633\pm0.005$  & $2.220\pm0.004$ & $1.322\pm0.003$ & $4.110\pm0.035$ \\
$V$     & 0.5519 & $0.910\pm0.003$  & $1.680\pm0.002 $& 1           & $3.110\pm0.026$ \\
$g$     & 0.4925 & 1$.280\pm0.000$  & $1.956\pm0.000$ & $1.164\pm0.002$ & $3.621\pm0.030$ \\
$r$     & 0.6205 & $0.555\pm0.000$  & $1.415\pm0.000$ & $0.842\pm0.001$ & $2.619\pm0.022$ \\
$i$     & 0.7519 & $0.057\pm0.000$  & $1.043\pm0.000$ & $0.621\pm0.001$ & $1.931\pm0.016$ \\
$z$     & 0.8665 & $-0.253\pm0.000$ & $0.811\pm0.000$ & $0.483\pm0.001$ & $1.502\pm0.013$ \\
$y$     & 0.9703 & $-0.442\pm0.000$ & $0.670\pm0.000$ & $0.399\pm0.001$ & $1.240\pm0.010$ \\
$J$     & 1.2340 & $-0.763\pm0.001$ & $0.430\pm0.001$ & $0.256\pm0.000$ & $0.796\pm0.007$ \\
$H$     & 1.6391 & $-0.993\pm0.001$ & $0.258\pm0.001$ & $0.154\pm0.000$ & $0.479\pm0.004$ \\
$\Ks$   & 2.1762 & $-1.114\pm0.001$ & $0.168\pm0.001$ & $0.100\pm0.000$ & $0.312\pm0.003$ \\
$W1$    & 3.3166 & $-1.192\pm0.001$ & $0.110\pm0.001$ & $0.065\pm0.001$ & $0.204\pm0.002$ \\
$W2$    & 4.5574 & $-1.229\pm0.001$ & $0.082\pm0.001$ & $0.049\pm0.001$ & $0.153\pm0.002$ \\
\noalign{\smallskip}\hline
\multicolumn{6}{c}{dwarfs with $-0.5 \leq {\rm[M/H]} < 0$ \& 5200\,K $\leq \Teff < 5700$\,K} \\
\hline 
NUV     & 0.2269 & $3.816\pm0.023$  & $3.975\pm0.018$ & $2.364\pm0.011$ & $6.878\pm0.039$ \\
$\GBP$  & 0.5232 & 1            & $1.780\pm0.000$ & $1.059\pm0.001$ & $3.079\pm0.011$ \\
$\GRP$  & 0.7649 & ...           & 1           & $0.595\pm0.000$ & $1.730\pm0.006$ \\
$G$     & 0.6180 & $0.515\pm0.000$  & $1.401\pm0.000$ & $0.833\pm0.001$ & $2.425\pm0.008$ \\
$B$     & 0.4451 & $1.615\pm0.002$  & $2.259\pm0.002$ & $1.344\pm0.001$ & $3.909\pm0.014$ \\
$V$     & 0.5503 & $0.874\pm0.002$  & $1.681\pm0.001$ & 1           & $2.909\pm0.010$ \\
$g$     & 0.4897 & $1.227\pm0.000$  & $1.956\pm0.000$ & $1.164\pm0.001$ & $3.385\pm0.012$ \\
$r$     & 0.6199 & $0.523\pm0.000$  & $1.407\pm0.000$ & $0.837\pm0.001$ & $2.435\pm0.009$ \\
$i$     & 0.7514 & $0.042\pm0.000$  & $1.032\pm0.000$ & $0.614\pm0.000$ & $1.787\pm0.006$ \\
$z$     & 0.8661 & $-0.252\pm0.000$ & $0.803\pm0.000$ & $0.478\pm0.000$ & $1.390\pm0.005$ \\
$y$     & 0.9698 & $-0.435\pm0.000$ & $0.661\pm0.000$ & $0.393\pm0.000$ & $1.144\pm0.004$ \\
$J$     & 1.2328 & $-0.749\pm0.001$ & $0.416\pm0.000$ & 0$.247\pm0.000$ & $0.720\pm0.003$ \\
$H$     & 1.6385 & $-0.999\pm0.001$ & $0.221\pm0.001$ & $0.131\pm0.000$ & $0.382\pm0.002$ \\
$\Ks$   & 2.1761 & $-1.111\pm0.001$ & $0.134\pm0.001$ & $0.079\pm0.000$ & $0.231\pm0.001$ \\
$W1$    & 3.3166 & $-1.190\pm0.001$ & $0.072\pm0.001$ & $0.043\pm0.000$ & $0.125\pm0.001$ \\
$W2$    & 4.5600 & $-1.223\pm0.001$ & $0.047\pm0.001$ & $0.028\pm0.001$ & $0.080\pm0.002$ \\
\noalign{\smallskip}\hline
\multicolumn{6}{c}{dwarfs with $-0.5 \leq {\rm[M/H]} < 0$ \& 5700\,K $\leq \Teff < 6200$\,K} \\
\hline
NUV     & 0.2269 & 4$.066\pm0.012$  & $4.194\pm0.009 $& $2.510\pm0.006$ & $7.367\pm0.026$ \\
$\GBP$  & 0.5188 & 1            & $1.786\pm0.000$ & $1.068\pm0.001$ & $3.136\pm0.009$ \\
$\GRP$  & 0.7635 & ...           & 1           & $0.598\pm0.000$ & $1.756\pm0.005$ \\
$G$     & 0.6109 & $0.539\pm0.000$  & $1.423\pm0.000$ & $0.852\pm0.001$ & $2.500\pm0.007$ \\
$B$     & 0.4437 & $1.579\pm0.002$  & $2.241\pm0.001$ & $1.341\pm0.001$ & $3.936\pm0.011$ \\
$V$     & 0.5497 & $0.854\pm0.001$  & $1.671\pm0.001$ & 1           & $2.936\pm0.008$ \\
$g$     & 0.4885 & $1.198\pm0.000$  & $1.941\pm0.000$ & $1.161\pm0.001$ & $3.410\pm0.009$ \\
$r$     & 0.6193 & $0.510\pm0.001$  & $1.400\pm0.001$ & $0.838\pm0.001$ & $2.460\pm0.007$ \\
$i$     & 0.7511 & $0.026\pm0.001$  & $1.020\pm0.001$ & $0.611\pm0.001$ & $1.792\pm0.005$ \\
$z$     & 0.8660 & $-0.250\pm0.001$ & $0.804\pm0.000$ & $0.481\pm0.000$ & $1.412\pm0.004$ \\
$y$     & 0.9696 & $-0.424\pm0.000$ & $0.667\pm0.000$ & $0.399\pm0.000$ & $1.171\pm0.003$ \\
$J$     & 1.2323 & $-0.733\pm0.001$ & $0.424\pm0.000$ & $0.254\pm0.000$ & $0.745\pm0.002$ \\
$H$     & 1.6381 & $-0.978\pm0.001$ & $0.232\pm0.001$ & $0.139\pm0.000$ & $0.407\pm0.002$ \\
$\Ks$   & 2.1760 & $-1.087\pm0.001$ & $0.146\pm0.001$ & $0.087\pm0.000$ & $0.256\pm0.002$ \\
$W1$    & 3.3166 & $-1.164\pm0.001$ & $0.086\pm0.001$ & $0.051\pm0.001$ & $0.151\pm0.002$ \\
$W2$    & 4.5615 & $-1.194\pm0.001$ & $0.062\pm0.001$ & $0.037\pm0.001$ & $0.108\pm0.002$ \\
\noalign{\smallskip}\hline
\multicolumn{6}{c}{dwarfs with $-0.5 \leq {\rm[M/H]} < 0$ \& 6200\,K $\leq \Teff < 7000$\,K} \\
\hline 
FUV     & 0.1531 & $3.865\pm0.240$  & $4.162\pm0.197$ & $2.492\pm0.118$ & $6.996\pm0.331$ \\
NUV     & 0.2268 & $4.177\pm0.011$  & $4.417\pm0.009$ & $2.645\pm0.006$ & $7.425\pm0.024$ \\
$\GBP$  & 0.5103 & 1            & $1.818\pm0.000$ & $1.089\pm0.001$ & 3$.056\pm0.007$ \\
$\GRP$  & 0.7593 & ...           & 1           & $0.599\pm0.000$ & $1.681\pm0.004$ \\
$G$     & 0.5937 & $0.579\pm0.000$  & $1.473\pm0.000$ & $0.882\pm0.000$ & $2.476\pm0.006$ \\
$B$     & 0.4413 & $1.546\pm0.001$  & $2.265\pm0.001$ & $1.356\pm0.001$ & $3.807\pm0.009$ \\
$V$     & 0.5482 & $0.819\pm0.001$  & $1.670\pm0.001$ & 1           & $2.807\pm0.007$ \\
$g$     & 0.4858 & $1.168\pm0.001$  & $1.956\pm0.000$ & $1.171\pm0.001$ & $3.287\pm0.008$ \\
$r$     & 0.6178 & $0.487\pm0.001$  & $1.399\pm0.001$ & $0.838\pm0.001$ & $2.351\pm0.006$ \\
$i$     & 0.7502 & $0.023\pm0.001$  & $1.019\pm0.001$ & $0.610\pm0.001$ & $1.713\pm0.004$ \\
$z$     & 0.8656 & $-0.257\pm0.001$ & $0.790\pm0.001$ & $0.473\pm0.000$ & $1.328\pm0.003$ \\
$y$     & 0.9693 & $-0.423\pm0.000$ & $0.654\pm0.000$ & $0.392\pm0.000$ & $1.100\pm0.003$ \\
$J$     & 1.2308 & $-0.727\pm0.000$ & $0.405\pm0.000$ & $0.243\pm0.000$ & $0.681\pm0.002$ \\
$H$     & 1.6374 & $-0.984\pm0.001$ & $0.195\pm0.001$ & $0.117\pm0.000 $& $0.328\pm0.001$ \\
$\Ks$   & 2.1759 & $-1.087\pm0.001$ & $0.111\pm0.001$ & $0.066\pm0.000$ & $0.186\pm0.001$ \\
$W1$    & 3.3166 & $-1.156\pm0.001$ & $0.054\pm0.001$ & $0.032\pm0.000$ & $0.091\pm0.001$ \\
$W2$    & 4.5641 & $-1.175\pm0.001$ & $0.039\pm0.001$ & $0.023\pm0.000$ & $0.066\pm0.001$ \\
\noalign{\smallskip}\hline
\multicolumn{6}{c}{dwarfs with $-0.5 \leq {\rm[M/H]} < 0$ \& 7000\,K $\leq \Teff \leq 8000$\,K} \\
\hline
FUV     & 0.1530 & $3.783\pm0.147$  & $4.150\pm0.122$ & $2.486\pm0.073$ & $7.067\pm0.211$ \\
NUV     & 0.2268 & $3.908\pm0.016$  & $4.254\pm0.014$ & $2.548\pm0.009$ & $6.532\pm0.038$ \\
$\GBP$  & 0.5045 & 1            & $1.833\pm0.000$ & $1.098\pm0.001$ & $3.121\pm0.014$ \\
$\GRP$  & 0.7563 & ...           & 1           & $0.599\pm0.001$ & $1.703\pm0.008$ \\
$G$     & 0.5799 & $0.608\pm0.000$  & $1.506\pm0.000$ & $0.902\pm0.001$ & $2.565\pm0.012$ \\
$B$     & 0.4403 & $1.509\pm0.002$  & $2.257\pm0.002$ & $1.352\pm0.002$ & $3.843\pm0.018$ \\
$V$     & 0.5467 & $0.804\pm0.002$  & $1.670\pm0.002$ & 1           & $2.843\pm0.013$ \\
$J$     & 1.2296 & $-0.721\pm0.001$ & $0.400\pm0.001$ & $0.239\pm0.000$ & $0.680\pm0.003$ \\
$H$     & 1.6371 & $-0.994\pm0.001$ & $0.173\pm0.001$ & $0.103\pm0.001$ & $0.294\pm0.002$ \\
$\Ks$   & 2.1758 & $-1.091\pm0.001$ & $0.091\pm0.001$ & $0.055\pm0.001$ & $0.156\pm0.002$ \\
$W1$    & 3.3164 & $-1.153\pm0.002$ & $0.040\pm0.001$ & $0.024\pm0.001$ & $0.068\pm0.002$ \\
$W2$    & 4.5645 & $-1.175\pm0.002$ & $0.022\pm0.001$ & $0.013\pm0.001$ & $0.037\pm0.002$ \\
\noalign{\smallskip}\hline
\multicolumn{6}{c}{dwarfs with $0 \leq {\rm[M/H]} \leq 0.5$ \& 4200\,K $\leq \Teff < 4700$\,K}  \\
\hline 
$\GBP$ & 0.5476 & 1            & $1.707\pm0.000$ & $1.001\pm0.003$ & $3.220\pm0.076$ \\
$\GRP$ & 0.7738 & ...           & 1           & $0.586\pm0.002$ & $1.886\pm0.044$ \\
$G$    & 0.6558 & $0.401\pm0.000$  & $1.284\pm0.000$ & $0.752\pm0.002$ & 2$.421\pm0.057$ \\
$B$    & 0.4537 & $1.748\pm0.016$  & $2.236\pm0.011$ & $1.311\pm0.008$ & $4.218\pm0.101$ \\
$V$    & 0.5563 & $0.998\pm0.007$  & $1.706\pm0.005$ & 1           & $3.218\pm0.076$ \\
$g$    & 0.4964 & $1.363\pm0.001$  & $1.964\pm0.000$ & $1.151\pm0.003$ & $3.704\pm0.087$ \\
$r$    & 0.6222 & $0.621\pm0.000$  & $1.439\pm0.000$ & $0.843\pm0.002$ & $2.714\pm0.064$ \\
$i$    & 0.7533 & $0.080\pm0.000$  & $1.056\pm0.000$ & $0.619\pm0.002$ & $1.992\pm0.047$ \\
$z$    & 0.8672 & $-0.239\pm0.000$ & $0.831\pm0.000$ & $0.487\pm0.001$ & $1.567\pm0.037$ \\
$y$    & 0.9708 & $-0.430\pm0.000$ & $0.696\pm0.000$ & $0.408\pm0.001$ & $1.313\pm0.031$ \\
$J$    & 1.2348 & $-0.748\pm0.001$ & $0.471\pm0.001$ & $0.276\pm0.001$ & $0.888\pm0.021$ \\
$H$    & 1.6400 & $-0.984\pm0.00$1 & $0.304\pm0.001$ & $0.178\pm0.001$ & $0.573\pm0.014$ \\
$\Ks$  & 2.1758 & $-1.101\pm0.002$ & $0.221\pm0.001$ & $0.130\pm0.001$ & $0.418\pm0.010$ \\
$W1$   & 3.3173 & $-1.192\pm0.002$ & $0.157\pm0.001$ & $0.092\pm0.000$ & $0.296\pm0.007$ \\
$W2$   & 4.5518 & $-1.249\pm0.002$ & $0.117\pm0.001$ & $0.068\pm0.001$ & $0.220\pm0.006$ \\
\noalign{\smallskip}\hline
\multicolumn{6}{c}{dwarfs with $0 \leq {\rm[M/H]} \leq 0.5$ \& 4700\,K $\leq \Teff < 5200$\,K}  \\
\hline
NUV     & 0.2268 & $3.877\pm0.168$  & $3.861\pm0.124$ & $2.296\pm0.074$ & $7.070\pm0.252$ \\
$\GBP$  & 0.5351 & 1            & $1.738\pm0.000$ & $1.033\pm0.003$ & $3.182\pm0.048$ \\
$\GRP$  & 0.7674 & ...           & 1           & $0.595\pm0.002$ & $1.831\pm0.028$ \\
$G$     & 0.6334 & $0.469\pm0.000$  & $1.346\pm0.000$ & $0.801\pm0.002$ & $2.465\pm0.038$ \\
$B$     & 0.4497 & $1.664\pm0.010$  & $2.228\pm0.007$ & $1.325\pm0.005$ & $4.079\pm0.064$ \\
$V$     & 0.5524 & $0.924\pm0.006$  & $1.682\pm0.004$ & 1           & $3.079\pm0.048$ \\
$g$     & 0.4931 & $1.294\pm0.001$  & $1.955\pm0.001$ & $1.162\pm0.003$ & $3.580\pm0.055$ \\
$r$     & 0.6207 & $0.567\pm0.001$  & $1.418\pm0.001$ & $0.843\pm0.002$ & $2.597\pm0.040$ \\
$i$     & 0.7519 & $0.064\pm0.001$  & $1.047\pm0.001$ & $0.623\pm0.002$ & $1.918\pm0.029$ \\
$z$     & 0.8666 & $-0.247\pm0.001$ & $0.818\pm0.001$ & $0.486\pm0.001$ & $1.497\pm0.023$ \\
$y$     & 0.9703 & $-0.437\pm0.001$ & $0.678\pm0.001$ & $0.403\pm0.001$ & $1.241\pm0.019$ \\
$J$     & 1.2340 & $-0.769\pm0.002$ & $0.432\pm0.001$ & $0.257\pm0.001$ & $0.791\pm0.012$ \\
$H$     & 1.6391 & $-1.004\pm0.002$ & $0.259\pm0.002$ & $0.154\pm0.001$ & $0.475\pm0.008$ \\
$\Ks$   & 2.1759 & $-1.123\pm0.002$ & $0.171\pm0.002$ & $0.102\pm0.001$ & $0.314\pm0.006$ \\
$W1$    & 3.3163 & $-1.213\pm0.003$ & $0.105\pm0.002$ & $0.062\pm0.001$ & $0.192\pm0.005$ \\
$W2$    & 4.5551 & $-1.251\pm0.003$ & $0.077\pm0.002$ & $0.046\pm0.001$ & $0.141\pm0.005$ \\
\noalign{\smallskip}\hline
\multicolumn{6}{c}{dwarfs with $0 \leq {\rm[M/H]} \leq 0.5$ \& 5200\,K $\leq \Teff < 5700$\,K}  \\
\hline 
NUV     & 0.2268 & $3.940\pm0.069$  & $4.037\pm0.053$ & $2.401\pm0.032$ & $7.102\pm0.121$ \\
$\GBP$  & 0.5257 & 1            & $1.771\pm0.000$ & $1.053\pm0.003$ & $3.115\pm0.034$ \\
$\GRP$  & 0.7647 & ...           & 1           & $0.595\pm0.001$ & $1.759\pm0.019$ \\
$G$     & 0.6202 & $0.511\pm0.000$  & $1.394\pm0.000$ & $0.829\pm0.002$ & $2.453\pm0.027$ \\
$B$     & 0.4461 & $1.621\pm0.006$  & $2.250\pm0.005$ & $1.338\pm0.004$ & $3.958\pm0.044$ \\
$V$     & 0.5506 & $0.884\pm0.005$  & $1.681\pm0.004$ & 1           & $2.958\pm0.033$ \\
$g$     & 0.4902 & $1.241\pm0.001$  & $1.956\pm0.001$ & $1.163\pm0.003$ & $3.441\pm0.038$ \\
$r$     & 0.6200 & $0.533\pm0.001$  & $1.411\pm0.001$ & $0.839\pm0.002$ & $2.482\pm0.027$ \\
$i$     & 0.7514 & $0.039\pm0.003$  & $1.030\pm0.002$ & $0.613\pm0.002$ & $1.812\pm0.020$ \\
$z$     & 0.8662 & $-0.254\pm0.001$ & $0.805\pm0.001$ & $0.478\pm0.001$ & $1.415\pm0.016$ \\
$y$     & 0.9698 & $-0.434\pm0.002$ & $0.666\pm0.001$ & $0.396\pm0.001$ & $1.171\pm0.013$ \\
$J$     & 1.2328 & $-0.759\pm0.002$ & $0.415\pm0.001$ & $0.247\pm0.001$ & $0.730\pm0.008$ \\
$H$     & 1.6386 & $-1.012\pm0.003$ & $0.220\pm0.002$ & $0.131\pm0.001$ & $0.387\pm0.006$ \\
$\Ks$   & 2.1759 & $-1.120\pm0.003$ & $0.137\pm0.002$ & $0.081\pm0.001$ & $0.241\pm0.005$ \\
$W1$    & 3.3162 & $-1.199\pm0.003$ & $0.076\pm0.003$ & $0.045\pm0.002$ & $0.134\pm0.005$ \\
$W2$    & 4.5581 & $-1.234\pm0.004$ & $0.049\pm0.003$ & $0.029\pm0.002$ & $0.086\pm0.005$ \\
\noalign{\smallskip}\hline
\multicolumn{6}{c}{dwarfs with $0 \leq {\rm[M/H]} \leq 0.5$ \& 5700\,K $\leq \Teff < 7000$\,K}  \\
\hline 
NUV     & 0.2269 & $4.023\pm0.044$  & $4.196\pm0.035$ & $2.523\pm0.022$ & $7.227\pm0.100$ \\
$\GBP$  & 0.5178 & 1            & $1.794\pm0.000$ & $1.079\pm0.002$ & $3.091\pm0.034$ \\
$\GRP$  & 0.7619 & ...           & 1           & $0.601\pm0.001$ & $1.722\pm0.019$ \\
$G$     & 0.6067 & $0.562\pm0.001$  & $1.447\pm0.001$ & $0.870\pm0.002$ & $2.492\pm0.028$ \\
$B$     & 0.4437 & $1.566\pm0.007$  & $2.244\pm0.005$ & $1.349\pm0.004$ & $3.864\pm0.044$ \\
$V$     & 0.5493 & $0.835\pm0.005$  & $1.663\pm0.004$ & 1           & $2.864\pm0.032$ \\
$g$     & 0.4881 & $1.216\pm0.004$  & $1.966\pm0.003$ & $1.182\pm0.003$ & $3.386\pm0.038$ \\
$r$     & 0.6188 & $0.513\pm0.005$  & $1.407\pm0.004$ & $0.846\pm0.003$ & $2.424\pm0.028$ \\
$i$     & 0.7508 & $0.013\pm0.007$  & $1.010\pm0.006$ & $0.607\pm0.004$ & $1.740\pm0.022$ \\
$z$     & 0.8659 & $-0.253\pm0.005$ & $0.799\pm0.004$ & $0.481\pm0.003$ & $1.377\pm0.017$ \\
$y$     & 0.9696 & $-0.429\pm0.003$ & $0.659\pm0.003$ & $0.396\pm0.002$ & $1.136\pm0.013$ \\
$J$     & 1.2321 & $-0.746\pm0.002$ & $0.407\pm0.001$ & $0.245\pm0.002$ & $0.702\pm0.008$ \\
$H$     & 1.6379 & $-1.014\pm0.003$ & $0.195\pm0.003$ & $0.117\pm0.002$ & $0.335\pm0.006$ \\
$\Ks$   & 2.1760 & $-1.120\pm0.003$ & $0.110\pm0.002$ & $0.066\pm0.002$ & $0.189\pm0.005$ \\
$W1$    & 3.3160 & $-1.180\pm0.004$ & $0.063\pm0.003$ & $0.038\pm0.002$ & $0.108\pm0.005$ \\
$W2$    & 4.5612 & $-1.199\pm0.004$ & $0.048\pm0.003$ & $0.029\pm0.002$ & $0.082\pm0.005$ \\
\noalign{\smallskip}\hline
\multicolumn{6}{c}{giants with   $-0.5\leq {\rm[M/H]} \leq 0.2$ \& 4200\,K $\leq \Teff \leq 5200$\,K}  \\
\hline 
$\GBP$ & 0.5387  & 1            & $1.700\pm0.000$ & $1.025\pm0.001$ & $2.893\pm0.015$ \\
$\GRP$ & 0.7667  & ...           & 1           & $0.603\pm0.001$ & $1.702\pm0.009$ \\
$G$    & 0.6419  & $0.437\pm0.000$  & $1.306\pm0.000$ & $0.787\pm0.001$ & $2.222\pm0.011$ \\
$B$    & 0.4525  & $1.781\pm0.003$  & $2.247\pm0.002$ & $1.354\pm0.002$ & $3.823\pm0.020$ \\
$V$    & 0.5525  & $0.941\pm0.003$  & $1.659\pm0.002$ & 1           & $2.823\pm0.015$ \\
$J$    & 1.2345  & $-0.814\pm0.002$ & $0.430\pm0.001$ & $0.259\pm0.001$ & $0.732\pm0.004$ \\
$H$    & 1.6393  & $-1.109\pm0.003$ & $0.224\pm0.002$ & $0.135\pm0.001$ & $0.381\pm0.004$ \\
$\Ks$  & 2.1757  & $-1.229\pm0.003$ & $0.140\pm0.002$ & $0.084\pm0.001$ & $0.238\pm0.003$ \\
$W1$   & 3.3172  & $-1.318\pm0.003$ & $0.077\pm0.002$ & $0.047\pm0.001$ & $0.132\pm0.003$ \\
$W2$   & 4.5501  & $-1.330\pm0.003$ & $0.069\pm0.002$ & $0.041\pm0.001$ & $0.117\pm0.004$ \\
$W3$   & 11.7281 & $-1.335\pm0.003$ & $0.065\pm0.002$ & $0.039\pm0.001$ & $0.111\pm0.003$ \\
\noalign{\smallskip}\hline
\end{longtable}
\end{center}

\end{document}